\PassOptionsToPackage{final}{graphicx}
\documentclass[final,a4paper,UKenglish,cleveref,numberwithinsect,autoref]{lipics-v2021}
\pdfoutput=1 %

\nolinenumbers

\hideLIPIcs %

\newboolean{spellingmode}
\setboolean{spellingmode}{false}

\newboolean{proofsinappendix}
\setboolean{proofsinappendix}{true}

\usepackage{todonotes}
\usepackage[marginclue,nomargin,footnote]{fixme}
\usepackage{stmaryrd}
\usepackage{xspace}
\usepackage{ifdraft}
\usepackage{tikz}
\usepackage{environ}
\usepackage{newfile}
\usepackage{mathtools}
\usepackage{hyperref}

\ifthenelse{\boolean{spellingmode}}{%
	\nolinenumbers
	\AtEndPreamble{
	\hyphenpenalty=50000 %
	\pagestyle{empty}%
	\thispagestyle{empty}%
	\def\pagestyle#1{}%
	\def\thispagestyle#1{}%
	\def\labelmarginpar#1{}%
	}
}{}

\newcommand{\takeout}[1]{}

\newcommand{\resetCurThmBraces}{%
  \gdef\curThmBraceOpen{(}%
  \gdef\curThmBraceClose{)}}
\resetCurThmBraces
\newcommand{\removeThmBraces}{%
  \gdef\curThmBraceOpen{}%
  \gdef\curThmBraceClose{}}
\resetCurThmBraces

\newenvironment{notheorembrackets}{\removeThmBraces}{\resetCurThmBraces}

\usepackage{etoolbox}
\patchcmd{\thmhead}{(#3)}{\curThmBraceOpen #3\curThmBraceClose }{}{}

\makeatletter
\ifthenelse{\boolean{proofsinappendix}}{
  \newoutputstream{proofstream}
  \openoutputfile{\jobname-proofs.out}{proofstream}
}{
}

\newcommand{\proofappendixbegin}[2]{%
  \phantomsection%
  \subsection*{\textbf{#1~\autoref{#2}}}%
  \addcontentsline{toc}{subsection}{#1~\autoref{#2}}%
  \label{#2:proof}%
  \def\proofappendix@qedsymbolmissing{\qed}
}
\newcommand{\proofappendixend}{%
  \proofappendix@qedsymbolmissing%
}
\let\oldqedhere\qedhere
\def\qedhere{\global\def\proofappendix@qedsymbolmissing{}\oldqedhere}

\newenvironment{proofhere}[2][Proof of]{%
  \subsection*{#1~\autoref{#2}.}%
  \def\proofappendix@qedsymbolmissing{\qed}%
}{%
  \proofappendixend%
  \par%
}

\ifthenelse{\boolean{proofsinappendix}}{%
  \NewEnviron{proofappendix}[2][Proof of]{%
    \addtostream{proofstream}{\noexpand\proofappendixbegin{#1}{#2}}%
    \addtostream{proofstream}{\noexpand\takeout{}}
    \expandafter\addtostream{proofstream}{\expandonce{\BODY}}%
    \addtostream{proofstream}{\noexpand\proofappendixend}%
    \addtostream{proofstream}{}%
  }%
}{%
  \newenvironment{proofappendix}[2][Proof of]{%
    \begin{proofhere}[#1]{#2}
  }{%
    \end{proofhere}
  }
}
\makeatother

\hypersetup{
  final,
  hidelinks}

\usetikzlibrary{cd,decorations,fit,arrows,automata}

\FXRegisterAuthor{tw}{twm}{TW}%
\FXRegisterAuthor{jr}{jrm}{JR}%
\title{Bisimilar States in Uncertain Structures}
\author{Jurriaan Rot}{Radboud University, Nijmegen, the
Netherlands\and\url{https://jurriaan.me}}{j.rot@cs.ru.nl}{}{}
\author{Thorsten Wißmann}{Radboud University, Nijmegen, the Netherlands\and
Friedrich-Alexander-Universität Erlangen-Nürnberg,
Germany\and\url{https://thorsten-wissmann.de}}{thorsten.wissmann@fau.de}{https://orcid.org/0000-0001-8993-6486}{}
\authorrunning{J.\ Rot and T.\ Wißmann}
\Copyright{J.\ Rot and T.\ Wißmann}

\newcommand{\lsharp}{\ensuremath{{L^\#}}}
\newcommand{\textqt}[1]{`#1'}
\newcommand{\Set}{\ensuremath{\mathsf{Set}}}
\newcommand{\id}{\ensuremath{\mathsf{id}}}
\newcommand{\M}{\ensuremath{\mathcal{M}}\xspace}
\newcommand{\SA}{\ensuremath{\mathcal{S}}\xspace}
\def\S{\SA}
\newcommand{\mealyN}{\ensuremath{\mathcal{N}}\xspace}
\newcommand{\C}{\ensuremath{\mathcal{C}}}
\newcommand{\Eq}{\ensuremath{\mathsf{Eq}}}
\newcommand{\Rel}{\ensuremath{\mathsf{Rel}}}
\newcommand{\Pos}{\ensuremath{\mathsf{Pos}}}
\newcommand{\inl}{\ensuremath{\mathsf{inl}}}
\newcommand{\xtransto}[1]{\xrightarrow{#1}}

\newcommand{\Powf}{\ensuremath{\mathcal{P}_{\mathsf{f}}}}
\newcommand{\anysymbolneqq}[1]{\mathrel{%
  \begin{tikzpicture}[baseline=(base)]
    \node[overlay] (neq) {\ensuremath{=}};
    \def\strikeOutX{2pt}
    \def\strikeOutY{2pt}
    \coordinate (strike out center) at ([yshift=.5pt]neq.center);
    \draw[overlay] ([xshift=\strikeOutX,yshift=\strikeOutY]strike out center)
    -- ([xshift=-\strikeOutX,yshift=-\strikeOutY]strike out center);
    \coordinate (base) at ([yshift=2pt]neq.center);
    \node[inner sep=0pt] (mainsymbol) at ([yshift=1pt]neq.north) {\ensuremath{#1}};
  \end{tikzpicture}
}}
\newcommand{\sqsubsetneqq}{\anysymbolneqq{\sqsubset}}
\newcommand{\sqsupsetneqq}{\anysymbolneqq{\sqsupset}}
\newcommand{\inr}{\ensuremath{\mathsf{inr}}}
\newcommand{\uncertOP}{\ensuremath{%
  \hat F_{\sqsubseteq}%
}}
\newcommand{\Coalg}{\ensuremath{\mathsf{Coalg}}}

\newcommand{\set}[2][]{%
  \ifthenelse{\equal{#2}{}}{%
    \ensuremath{{#1\emptyset}}%
  }{%
    \ensuremath{{#1\{#2#1\}}}%
  }%
}

\newcommand{\N}{\ensuremath{\mathbb{N}}}

\newcommand{\fpair}[1]{{\langle #1\rangle}}
\newcommand{\nonempty}{\mathsf{ne}}
\newcommand{\partialto}{\mathbin{\rightharpoonup}}
\newcommand{\nepartialto}{\mathbin{{\rightharpoonup}{\mathllap{\raisebox{-2pt}{\ensuremath{\scriptstyle\nonempty}}}}}}

\newcommand{\commutes}{\ensuremath{\circlearrowleft}}
\newcommand{\laxto}{\mathrel{%
\begin{tikzcd}[]%
  |[minimum width=0pt,inner sep=0pt]| {}
    \arrow{r}[description,inner xsep=0pt,pos=0.45]{\,\sqsubseteq} \pgfmatrixnextcell
  |[minimum width=0pt,inner sep=0pt]| {}
\end{tikzcd}%
}}
\newcommand{\oplaxto}{\mathrel{%
\begin{tikzcd}[]%
  |[minimum width=0pt,inner sep=0pt]| {}
    \arrow{r}[description,inner xsep=0pt,pos=0.45]{\,\sqsupseteq} \pgfmatrixnextcell
  |[minimum width=0pt,inner sep=0pt]| {}
\end{tikzcd}%
}}
\newcommand{\op}{\mathsf{op}}
\newcommand{\question}{\ensuremath{\mathord{\text{\upshape\sffamily\bfseries
?}}}}

\newcommand{\descto}[3][]{\arrow[phantom]{#2}[#1]{\text{\footnotesize{}\begin{tabular}{c}#3\end{tabular}}}}
\newcommand{\semantics}[2][]{\ensuremath{#1\llbracket #2 #1\rrbracket}}

\tikzset{shiftarr/.style={
        rounded corners,%
        to path={--([#1]\tikztostart.center)
                     -- ([#1]\tikztotarget.center) \tikztonodes
                     -- (\tikztotarget)},
}}

\newtheorem{assumption}[theorem]{Assumption}
\newtheorem{notation}[theorem]{Notation}

\tikzset{
  loop at/.style={
    loop,
    out=#1-30,
    in=#1+30,
    looseness=6,
    every node/.append style={
      anchor=#1-180,
    },
  },
}

\tikzset{
    longmapsto/.style={
      preaction={
        draw,
        -,
        draw=white,
        line width=4pt,
      },
      /tikz/commutative diagrams/every arrow,
      /tikz/commutative diagrams/every label,
      shorten <= 4pt,
      shorten >= 4pt,
      bend angle=10,
      /tikz/commutative diagrams/mapsto,
    },
}

\tikzset{
  automaton/.style={
    initial text={},
    transition/.style={
      auto,
      every node/.append style={
        every label,
      },
    },
    every edge/.append style={
      ->,
      >={Straight Barb[scale=0.8]},
      thick,
      draw=lipicsLineGray,
      line width=1pt,
      shorten <= 2pt,
      shorten >= 2pt,
      bend angle=20,
    },
    every label/.append style={
      align=center,
      text=black,
    },
    state/.append style={
      inner sep=1pt,
      minimum size=20pt,
      text depth=0pt,
      text = black,
      draw = none,
      fill = lipicsYellow,
    },
  },
  coalgebra frame/.style={
    draw=lipicsLineGray,
    rounded corners=4pt,
    line width=1pt,
    inner sep = 4mm,
  },
  coalgebra name/.style={
    outer sep=5pt,
    inner sep=2pt,
    rounded corners=1pt,
    fill=lipicsLineGray,
    text=white,
  },
}


\category{} %

\begin{document}
\maketitle

\thispagestyle{empty}
\begin{abstract}
We provide a categorical notion called uncertain bisimilarity, which allows to
reason about bisimilarity in combination with a lack of knowledge about the
involved systems. Such uncertainty arises naturally in automata learning
algorithms, where one investigates whether two observed behaviours come from the same
internal state of a black-box system that can not be transparently inspected.
We model this uncertainty as a set functor equipped with a partial order which
describes possible future developments of the learning game. On such a functor,
we provide a lifting-based definition of uncertain bisimilarity and verify
basic properties. Beside its applications to Mealy machines, a natural model for
automata learning, our framework also instantiates to an existing compatibility 
relation on suspension automata, which are used in model-based testing.
We show that uncertain bisimilarity is a necessary but not sufficient condition
for two states being implementable by the same state in the black-box system.
To remedy the failure of the one direction, we characterize uncertain
bisimilarity in terms of coalgebraic simulations.
\end{abstract}

\section{Introduction}
Inspired from constructive mathematics, Geuvers and Jacobs~\cite{GJ2021}
introduced apartness relations on coalgebraic systems, complementing
bisimilarity. While bisimilarity is a coinductive characterization of behavioural equivalence, 
apartness is inductive, and allows constructing \emph{finite} proofs of difference in behaviour. 

Although apartness
and bisimilarity are just different sides of the same coin, the angle of
\textqt{apartness} turned out to be fruitful in the recent $\lsharp$ automata learning
algorithm~\cite{VGRW22}.
This algorithm works in the active learning setting of Angluin~\cite{Angluin87}, where
a \emph{learner} tries to reconstruct the implementation of an automaton (or concretely a Mealy machine
in~\cite{VGRW22}) from only its black-box behaviour. In $\lsharp$, a crucial task of
the learner is to determine whether two input words $w$, $v$ lead to the
identical or to distinct states in the black box. Throughout the learning game,
the learner makes more and more observations. If at some point
the learner finds out that the states $q_w$, $q_v$ reached by $w$ and $v$
respectively have different behaviours, then $q_w$ and $q_v$ are provably
different -- called \emph{apart}. For that, it is not required that we know the
entire semantics of $q_w$ and $q_v$; instead, it suffices to observe one aspect
of their behaviour in which they differ in incompatible ways.

Once states turn out to be apart, they stay
so throughout the entire remaining learning game, no matter which further
observations of the black box are made. Thus, the apartness relation grows
monotonically as the learning game progresses.
This beauty of monotonicity breaks if we consider
\emph{bisimilarity}: as long as states $q_w$,
$q_v$ have not been proven different yet, should they be considered bisimilar?
Or do we just have insufficiently much information at hand?
If we do not know the number of states in the black box,
we can never consider states $q_w$, $q_v$ bisimilar with 100\% certainty during
the learning process.

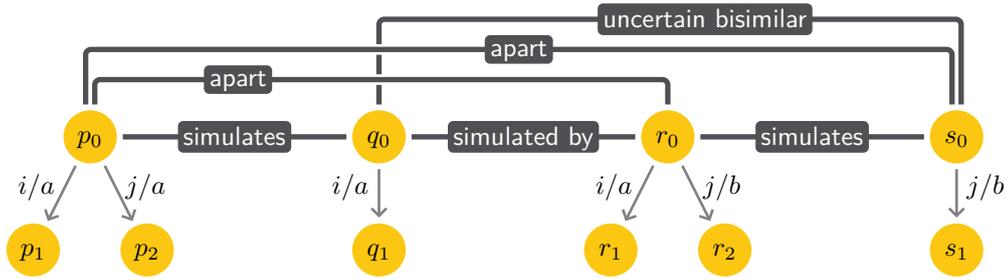
\begin{figure}[t]
  \centering
  \begin{tikzpicture}[x=15mm,y=15mm,
      automaton/.append style={
        transition/.append style={
          every node/.append style={
            pos=0.4, %
          },
        },
      },
    ]
    \def\subautomatondistance{3.8cm}
    \begin{scope}[automaton,xshift=-\subautomatondistance]
      \def\stateFamilyName{p}
      \node[state] (p0) {$\stateFamilyName_0$};
      \node[state] (p1) at (-0.5,-1) {$\stateFamilyName_1$};
      \node[state] (p2) at ( 0.5,-1) {$\stateFamilyName_2$};
      \path[transition] (p0) edge node[left] {$i/a$} (p1);
      \path[transition] (p0) edge node[right] {$j/a$} (p2);
    \end{scope}
    \begin{scope}[automaton,xshift=0cm]
      \def\stateFamilyName{q}
      \node[state] (q0) {$\stateFamilyName_0$};
      \node[state] (q1) at (0,-1) {$\stateFamilyName_1$};
      \path[transition] (q0) edge node[swap] {$i/a$} (q1);
    \end{scope}
    \begin{scope}[automaton,xshift=\subautomatondistance]
      \def\stateFamilyName{r}
      \node[state] (r0) {$\stateFamilyName_0$};
      \node[state] (r1) at (-0.5,-1) {$\stateFamilyName_1$};
      \node[state] (r2) at ( 0.5,-1) {$\stateFamilyName_2$};
      \path[transition] (r0) edge node[left] {$i/a$} (r1);
      \path[transition] (r0) edge node[right] {$j/b$} (r2);
    \end{scope}
    \begin{scope}[automaton,xshift=2*\subautomatondistance]
      \def\stateFamilyName{s}
      \node[state] (s0) {$\stateFamilyName_0$};
      \node[state] (s1) at (0,-1) {$\stateFamilyName_1$};
      \path[transition] (s0) edge node[right] {$j/b$} (s1);
    \end{scope}
    \begin{scope}[rel/.style={
      shorten <= 2pt,
      shorten >= 2pt,
      draw=lipicsGray,
      line width=2pt,
      rounded corners=2pt,
      preaction={
        draw,
        -,
        draw=white,
        line width=4pt,
      },
      every node/.append style={
        text=white,
        fill=lipicsGray,
        font=\sffamily,
        rounded corners=2pt,
        inner xsep=2pt,
        inner ysep=2pt,
        text depth=1pt,
        anchor=center,
      },
    }]
    \path[rel] (p0) to node {simulates} (q0);
    \path[rel] (q0) to node {simulated by} (r0);
    \path[rel] (r0) to node {simulates} (s0);
    \def\multiEdgeDistance{.6mm}
    \path[rel] (q0.north) -- +(0,12mm)
      -| node[pos=0.28] {uncertain bisimilar}
        ([xshift=\multiEdgeDistance]s0.north);
    \path[rel] ([xshift=-\multiEdgeDistance]p0.north) -- ++(0,8mm)
      -| node[pos=0.25] {apart} ([xshift=-\multiEdgeDistance]s0.north);
    \path[rel] ([xshift=\multiEdgeDistance]p0.north) -- +(0,4mm)
      -| node[pos=0.125] {apart} (r0.north);
    \end{scope}
  \end{tikzpicture}
  \caption{Partial Mealy machine for
  the input alphabet $\set{i,j}$ and output alphabet $\set{a,b}$.}
  \label{figNotionComparison}
\end{figure}

In the present work, we close this gap by introducing the notion of
\emph{uncertain bisimilarity}, which expresses that two states might be
bisimilar -- but we are not certain about it, because we simply did not observe any
reason yet that would disprove bisimilarity.
The main idea is exemplified by the Mealy machine in
\autoref{figNotionComparison}:
states $p_0$ and $r_0$ are apart, because
$p_0$ has output $a$ on input $j$ whereas $r_0$ yields a different output $b$ on
input $j$. By the same input $j$, we can tell that $p_0$ and $s_0$
are apart, even though we do not yet know the behaviour of $s_0$ on input $i$.
Furthermore, states $q_0$ and $s_0$ can either turn out to be
apart or to be bisimilar, depending on their outputs on input $j$. Thus, we call $q_0$ and $s_0$ \emph{uncertain
bisimilar}. If we for instance try to explore the output of $q_0$ on input
$j$, then depending on the output, $q_0$ will be identical to $p_0$
or $r_0$. Until we know this, $q_0$ is simulated by both
$p_0$ and by $r_0$. Simulation is a special case of uncertain bisimilarity,
because it not only says that the behaviours are compatible, but that one
behaviour is even included in the other.

Our aim is to establish a theory for
uncertain bisimilarity at the level of coalgebras, including the motivating
example of Mealy machines.
Working with bisimulation relations and bisimilarity benefits from a rich
categorical theory. In particular, they are themselves coalgebras in the
category of relations~\cite{HermidaJ98,jacobs-coalg}. Here, the coalgebraic type
functor considered on relations is a lifting of the original coalgebraic type
functor for the systems of interest. %

In the present paper, we incorporate the explicit treatment of the lack of
knowledge which is omnipresent in the learning setting.
Formally, we do this by equipping the type functor with a partial order. This order
$s\sqsubseteq t$ represents that the so far observed behaviour $s$ might be
extended to behaviour $t$ after additional observations. This order immediately
induces two further notions of lax coalgebra morphisms.

\subparagraph*{Contributions.}
With such a partial order on the type functor, we will:
\begin{itemize}
  \item Define a generic system equality notion, called \emph{uncertain
  bisimulation}, derived from the relation lifting of the type functor.

  \item We show basic properties of the relation such as reflexivity and
  symmetry. It is immediate that uncertain bisimilarity is not transitive, and
  thus, no equivalence relation.

  \item As instances, we discuss (partial) Mealy machines as the running
  example. Moreover, we cover suspension automata, for which uncertain
  bisimilarity instantiates to an existing compatibility notion, used in the \emph{ioco}
  conformance relation from model-based testing~\cite{BJM19}.

  \item It is a standard result that standard coalgebraic bisimilarity coincides with being
  identifiable by coalgebra morphisms (under often-met assumptions on the type
  functor)~\cite{Staton11}. We show that uncertain bisimilarity is not characterized via
  identifiability in lax coalgebra morphisms -- for the running example of
  Mealy machines.

  \item Instead, we show that uncertain bisimilarity is characterized via
  coalgebraic simulations -- for two definitions of
  coalgebraic simulation. Concretely, two states are uncertain bisimilar if
  there is a state in another coalgebra that simulates both (e.g.~$r_0$
  simulates both $q_0$ and $s_0$ in \autoref{figNotionComparison}). 
\end{itemize}
\twnote{}

\section{Preliminaries}
We first establish the basic coalgebraic notions used in the technical
development later, see e.g.~\cite{jacobs-coalg}. We assume that the reader is familiar with concepts from
basic category theory.
\begin{definition}
  For a functor $F\colon \C\to\C$, an \emph{$F$-coalgebra} $(C,c)$ is an object $C\in
  \C$ (the \emph{carrier}) together with a morphism $c\colon C\to FC$ in $\C$
  (the \emph{structure}). For two
  $F$-coalgebras, a coalgebra morphism $h\colon (C,c)\to (D,d)$ is a morphism
  $h\colon C\to D$ satisfying $Fh\cdot c= d\cdot h$.
  We denote the category of $F$-coalgebras by $\Coalg(F)$.
\end{definition}

Most of our coalgebras will live in the category $\Set$ of sets and maps and our
leading example is the functor modelling Mealy machines:
\begin{example}
\begin{enumerate}
\item For fixed sets $I$ and $O$ of input and output symbols, consider
the functor
\[
  \M_T\colon \Set\to\Set
  \qquad
  \M_T X = (O\times X)^I.
\]
An $\M_T$-coalgebra is then a set $C$ together with a map $c\colon C\to (O\times
C)^I$ which sends each state $q\in C$ and input symbol $i\in I$ to a pair of an
output symbol and a successor state to which the Mealy machine transitions:
$c(q)(i) \in O\times C$. We write
\[
  q\xrightarrow{i/o} q'
\]
to specify that $c(q)(i) = (o,q')$.
In the name of the functor, the index $T$ shall indicate that the Mealy machine
is \emph{total}, in the sense that it is defined for every input $i\in I$.

\item The finitary powerset functor $\Powf$ sends each set $X$ to the set of its
finite subsets $\Powf X=\set{S\subseteq X\mid S\text{ finite}}$
and maps $f\colon X\to Y$ to direct images:
\(
  \Powf f\colon \Powf X\to \Powf Y
\),
\(
  \Powf f(S) := \set{
    f(x) \mid x\in S
  }
\).

\end{enumerate}
\end{example}

A canonical domain for the semantics of coalgebras is the \emph{final}
coalgebra:
\begin{definition}
  The final $F$-coalgebra is the final object in $\Coalg(F)$. Concretely, a
  coalgebra $(D,d)$ is final if for every $(C,c)$ in $\Coalg(F)$ there is a
  unique coalgebra morphism $h\colon (C,c)\to (D,d)$. If it exists, we denote
  the final coalgebra for $F$ by $(\nu F, \tau)$ and the induced unique morphism
  for $(C,c)$ by $\semantics{-}\colon (C,c)\to (\nu F,\tau)$.
\end{definition}

\begin{example}\label{exFinalMealy}
  The final $\M_T$-coalgebra is carried by the set $\nu \M_T = O^{I^+}$ -- the
  set of all maps $I^+\to O$ from non-empty words $I^+$ to $O$.
\end{example}

Equivalently, we can characterize the semantics $\nu \M_T$ in terms of maps
$I^*\to O^*$ that interact nicely with the prefix-order on words:
\begin{notation}
  For words $v,w\in I^*$ (in particular also for non-empty words $I^+\subseteq
  I^*$), we write $v\le w$ to denote that $v$ is a prefix of $w$. The length of
  a word $w$ is denoted by $|w| \in \N$.
\end{notation}
Then, we can characterize $\nu \M_T$ 
as maps $I^*\to O^*$
that preserve length and prefixes of words:
\[
  \nu \M_T \cong \set{
    f\colon I^*\to O^* \mid
    \text{for all }w\in I^*\colon
    |f(w)| = |w|\text{ and }
    \text{for all }v\le w\colon
    f(v) \le f(w) }
\]

\section{A Lax Coalgebra Morphism Lacks Knowledge}
In the learning game for Mealy machines, the learner tries to reconstruct the
internal implementation of a Mealy machine
\[
  c\colon C\to \M_T C = (O\times C)^I
\]
by only its black-box behaviour. For that, one assumes
a distinguished initial state $q_0\in C$ and it is the task of the learner to
construct a Mealy machine with the same behaviour $\semantics{q_0}$ as that of $q_0$.
Being in a block-box setting means that the learner knows neither $C$ or $c$.
Instead, the learner can enter a word $i_1,\ldots,i_n\in I$ of input symbols
from the input alphabet $I$ to the black box, referred to as a \emph{query}, and then observe
the output symbols $o_1,\ldots,o_n \in O$. More precisely, the learner observes
the output symbols
\[
  o_k = \semantics{q_0}(\underbrace{i_1\cdots i_k}_{\in\,I^+}) \in O
  \qquad\text{for every }1\le k\le n
  \qquad\text{(with $\semantics{-}$ as in \autoref{exFinalMealy})}.
\]
On this query, the black box reveals the output $o_1\in O$ of the
initial state for input $i_1$. But after performing only this query, we still don't know the output for all the
other input symbols $i_1'\in I$, $i_1' \neq i_1$ with which we could have started the
input word.

After such a query, the black box returns to the initial state $q_0$ in
order to be ready for the next query. In concrete learning scenarios this reset
to initial state is for example realized by resetting the actual hardware of a
system that is learned. When learning network protocol implementations, this
reset-behaviour is realized by opening a separate network connection (or
session) for each new input query.

The $\lsharp$ algorithm (for Mealy machines) gathers all the information from
the performed input queries in an \emph{observation tree}. This bundles the
observations of the experiments so far in a single data structure. However, this
structure is not an $\M_T$-coalgebra itself, because the knowledge about the
outputs for some inputs $i\in I$ in some states in the tree will be
lacking.

We can model this lack of knowledge by the following functor
\[
  \M\colon \Set\to\Set
  \qquad
  \M X = (\set{\question} + O\times X)^I.
\]
The element \textqt{\question} models that we do not know the transition yet.
\begin{notation}
  We abbreviate partial functions via
  \(
    (X \partialto Y) := (\set{\question} + Y)^X.
  \)
\end{notation}

So we can also write $\M X = (I\partialto O\times X)$.
Compared to $\M_T$, a state $q$ in an $\M$-coalgebra $d\colon D\to \M D$ is
either undefined for an input $i\in I$, i.e.~$d(q)(i) = \question$, or has a
transition defined, i.e.~we have both an output $o\in O$ and a successor state
$q'\in D$ with
\[
  d(q)(i) = (o,q')
  \qquad\text{ or using notation: }\qquad
  q\xrightarrow{i/o} q'.
\]
This kind of partiality in $\M$ models that whenever the learner sends a word
$w\in I^+$, the black box reveals the output symbols of all transitions along
the way of processing $w$.
Thus, the semantics of states $q \in D$ in a such a partial Mealy machine,
i.e.~$\M$-coalgebras, can be characterized by:
\begin{equation}
  \nu \M := \set{
  f\colon I^+ \partialto O
  \mid
  \text{for all }
  v \le w
  \text{ if }
  f(w) \in O
  \text{ then }
  f(v) \in O
  }.
\label{nuPartialMealy}
\end{equation}
This monotonicity condition describes that whenever a learner has observed the
behaviour for an input word $w\in I^+$, then we also have observed the outputs
of all the prefixes $v\le w$.
\begin{proposition}\label{thmFinalPartialMealy}
  The final $\M$-coalgebra $(\nu\M,\tau)$ is characterized by
  \eqref{nuPartialMealy} and the map
  \[
    \tau\colon \nu \M \to (I\partialto O\times \nu \M)
    \qquad
    \tau(f) = ~~i\mapsto \begin{cases} 
      \question & \text{if }f(i) = \question \\
      (o,w\mapsto f(i\,w)) &\text{if }f(i) \in O
    \end{cases}
  \]
\end{proposition}
The structure sends every $f\in \nu\M$ to a successor structure of type
$\M\nu\M$. For $i\in I$, this successor structure yields $\tau(f)(i)\in
\set{\question} + O\times \nu \M$.
\begin{proofappendix}{thmFinalPartialMealy}
  For a coalgebra $c\colon C\to \M C$, we define a map
  \[
    \semantics{q}\colon I^+\to (\set{\question}+O)
  \]
  parametric in $q\in C$ by recursion on $w\in I^+$:
  \[
    \semantics{q}(i) = \M!(c(q)(i)) \in (\set{\question}+O\times 1) \cong
    (\set{\question}+O).
  \]
  \[
    \semantics{q}(i\,w) = \begin{cases}
      \semantics{p}(w) &\text{if }c(q)(i) = (o,p) \in O\times C\\
      \question &\text{if }c(q)(i) = \question
    \end{cases}.
  \]
  It is mechanical to check the antitonicity of $\semantics{q}$ and that
  $\semantics{-}$ is the unique coalgebra morphism.
\end{proofappendix}

During the learning process, the learner might be able to modify
the coalgebra after the output of state $q$ on input $i$ has been observed. In
this sense, we increase knowledge, and this can be modelled by the usual order
on partial functions:
\begin{definition}
For partial functions $t,s\colon A\partialto B$, we fix the partial order
\[
  t\sqsubseteq_{A\partialto B} s
  ~:\Longleftrightarrow~
  \forall i\in I\colon
  t(i) \in \set{s(i), \question}.
\]
The functor $\M X = (I\partialto O\times X)$ inherits the poset structure
$(\M X, \sqsubseteq)$ from partial maps.
\end{definition}
The equivalence means that for every input $i\in I$, the value of $t(i)$ is
either undefined ($t(i) = \question$) or agrees with $i$th entry in the other
successor structure ($t(i) = s(i)$). The partial order itself represents how the
behaviour can possibly be completed if we found out more information about the
full Mealy machines. That is, the partial order shows possible options in the
future learning process. 

This principle also works for other system types, so we generally assume (e.g.~\cite{HughesJ04}):
\begin{assumption}
  Fix a functor $F_\Pos\colon \Set\to \Pos$ and define:
  \begin{itemize}
    \item $F:= U\cdot F_\Pos$, where $U\colon \Pos\to \Set$ is the usual forgetful functor.
    \item Let $\sqsubseteq_{FX}$ be the order on $F_\Pos X$, i.e.~we have
    \( F_\Pos X = (FX,\sqsubseteq_{FX}) \).
  \end{itemize}
\end{assumption}

The functoriality of $F_\Pos$ means that for every $f\colon X\to Y$, the map
$Ff\colon FX\to FY$ is monotone. This partial order gives rise to a lax notion
of coalgebra morphisms:
\begin{definition} \label{laxMorphism}
  A \emph{lax $F$-coalgebra morphism} $h\colon (C,c)\laxto (D,d)$ between
  $F$-coalgebras is a map $h\colon C\to D$ such that for all $x\in C$ we have
  $Fh(c(x)) \sqsubseteq_{FD} d(h(x))$.
  \begin{figure}%
    \begin{minipage}{.48\textwidth}%
      \begin{tikzcd}
        C
        \arrow{r}{h}
        \arrow{d}[swap]{c}
        \descto{dr}{\(\sqsubseteq\)}
        & D
        \arrow{d}{d}
        \\
        FC
        \arrow{r}{Fh}
        & FD
      \end{tikzcd}
      \caption{Diagrammatic notation of a lax $F$-coalgebra morphism $h\colon
      (C,c)\laxto (D,d)$}
      \label{figDiagLaxMorphism}
    \end{minipage}%
    \hfill%
    \begin{minipage}{.48\textwidth}%
      \begin{tikzcd}
        C
        \arrow{r}{h}
        \arrow{d}[swap]{c}
        \descto{dr}{\(\sqsupseteq\)}
        & D
        \arrow{d}{d}
        \\
        FC
        \arrow{r}{Fh}
        & FD
      \end{tikzcd}
      \caption{Diagrammatic notation of an oplax $F$-coalgebra morphism $h\colon
      (C,c)\oplaxto (D,d)$}
      \label{figDiagOpLaxMorphism}
    \end{minipage}%
  \end{figure}
  We write $\sqsubseteq$ in squares to indicate lax commutativity as shown in
  \autoref{figDiagLaxMorphism}. Dually,
  an \emph{oplax $F$-coalgebra morphism} $h\colon (C,c)\oplaxto (D,d)$ is a map
  $h\colon C\to D$ such that for all $x\in C$, we have
  $Fh(c(x)) \sqsupseteq_{FD} d(h(x))$, and notated in diagrams as shown in
  \autoref{figDiagOpLaxMorphism}. In contrast,
  we write $\commutes$ to emphasize proper commutativity.
\end{definition}
Lax coalgebra morphisms could also be called functional simulations.
Intuitively, $h\colon (C,c)\laxto (D,d)$ means that $(D,d)$ has at
least as many transitions as $(C,c)$. Conversely, $h\colon (C,c)\oplaxto (D,d)$
means that $(C,c)$ has at least as many transitions as $(D,d)$.

In the learning game, lax coalgebra homomorphisms arise
naturally, because there, all observations are collected in an observation tree
$(T, t)$.
This observation tree is an $F$-coalgebra that admits a lax $F$-coalgebra
morphism $h\colon (T,t)\to (B,b)$ to the black-box $(B,b)$ that needs to be
learned. An example of lax morphisms for Mealy machines is visualized in \autoref{figExFunSim}.
\begin{figure}[t]\centering%
  \begin{tikzpicture}[x=2cm,y=2cm]
    \begin{scope}[automaton,xshift=-0.38*\textwidth]
      \node[state] (q0) {$q_0$};
      \node[state] (q1) at (0,-1) {$q_1$};
      \path[transition] (q0) edge node[right] {$i/o$} (q1);
      \node[coalgebra frame,fit=(q0) (q1)] (frame) {};
      \node[coalgebra name] at (frame.north) {$(T,t)$};
    \end{scope}
    \begin{scope}[automaton]
      \node[state] (p0) {$p_0$};
      \node[state] (p1) at (-0.5,-1) {$p_1$};
      \node[state] (p2) at (0.5,-1) {$p_1$};
      \path[transition] (p0) edge node[left] {$i/o$} (p1);
      \path[transition] (p0) edge node[right,xshift=1pt] {$j/o$} (p2);
      \node[coalgebra frame,fit=(p0) (p1) (p2)] (frame) {};
      \node[coalgebra name] at (frame.north) {$(T',t')$};
    \end{scope}
    \begin{scope}[automaton,xshift=0.38*\textwidth]
      \node[state] (r0)  {$r_0$};
      \node[state] (r1) at (0,-1) {$r_1$};
      \path[] (r0) edge node[right] {$i/o$} (r1);
      \path[] (r0) edge[loop at=0] node (r0 loop) {$j/o$} (r0);
      \node[coalgebra frame,fit={(r0) (r1) (r0 loop.center)}] (frame) {};
      \node[coalgebra name] at (frame.north) {$(B,b)$};
    \end{scope}
    \begin{scope}[
    ]
      \path[longmapsto,bend left] (q0) to node {$g$} (p0);
      \path[longmapsto,bend left] (q1) to node {$g$} (p1);
      \path[longmapsto,bend left] (p0) to node {$h$} (r0);
      \path[longmapsto,bend right=16] (p1) to node[pos=0.7] {$h$} (r1);
      \path[longmapsto,bend left] (p2) to node {$h$} (r0);
    \end{scope}
  \end{tikzpicture}
  \caption{Two lax $\M$-coalgebra morphism $g\colon (T,t)\laxto (T',t')$
  and $h\colon (T',t')\laxto (B,b)$ for $\M X = (\set{\question}+O\times X)^I$
  with $I=\set{i,j}$, $O=\set{o}$.}
  \label{figExFunSim}
\end{figure}

Of course, the learner only sees the observation tree $(T,t)$ but neither
$(B,b)$ nor $h$. But, the learner can make use of the fact that there is
\emph{some} lax coalgebra morphisms, and can use it to deduce properties of
$(B,b)$. The correctness proof of the \lsharp{} learning
algorithm~\cite{VGRW22} in fact relies on the existence of such a lax coalgebra
morphism.

\subparagraph{Conformance relations.}
Related to automata learning is the application of \emph{conformance testing} of
state-based systems. In particular, the \emph{ioco} (\textbf{i}nput \textbf{o}utput
\textbf{co}nformance) relation from testing theory~\cite{Tretmans08} nicely fits
into the coalgebraic theory too, while using a non-trivial order on the functor. The
systems of interest are \emph{suspension automata}, which are a subclass of
deterministic labelled transition systems and coalgebras for the following
functor:
\begin{definition}
  For partial maps which are defined on at least one input we write
  \[
    (A \nepartialto B)
    :=
    \set{f\colon A\partialto B\mid \exists a\in A\colon f(a) \neq \question}.
  \]
  For a fixed set of inputs $I$ and outputs $O$, define the \emph{suspension
  automaton} functor
  \[
    \SA X := (I\partialto X)\times (O\nepartialto X).
  \]
\end{definition}
Following existing presentations~\cite{BJM19,Tretmans08}, it is not hard to see
that suspension automata are coalgebras for this functor:
\begin{definition}
  A \emph{suspension automaton} is a finite $\SA$-coalgebra, i.e.~a finite set of
  states $C$ and a map $c\colon C\to \SA C$. For a given coalgebra $(C,c)$, we
  write $?$ for input transitions and $!$ for output transitions:
  \[
    x\xrightarrow{?a} y
    \quad :\Longleftrightarrow \quad
    \pi_1(c(x))(a) = y
  \qquad\text{and}\qquad
    x\xrightarrow{!a} y
    \quad :\Longleftrightarrow \quad
    \pi_2(c(x))(a) = y
  \]
\end{definition}
In other words, a suspension automaton is a deterministic LTS where the set of labels is partitioned
into inputs and outputs. For some of the
inputs and for some of the outputs, a suspension automaton in some state $x$ can
make a transition to another state. But every state is non-blocking in the sense
that for every state $x$ there is at least one output $o\in O$ such that $x$ can
make a transition $x\xrightarrow{!o} y$.

The conformance relation between specifications and implementations is related
to \emph{alternating} simulations. The detailed definition is not relevant now
(we will recall it later in \autoref{defIocoCompat}), because instead, we can
simply define this alternating flavour by reversing the order in the output part
of the functor $\SA$:
\begin{definition}\label{defOrderSA}
  For $(s_i,s_o)\in \SA X$ and $(t_i,t_o)\in \SA X$ we put:
  \[
    (s_i,s_o) \sqsubseteq_{\SA X} (t_i,t_o)
    \qquad:\Longleftrightarrow\qquad
    \underbrace{s_i \sqsubseteq t_i}_{\text{\upshape in $I\partialto X$}}
    \text{ and }
    \underbrace{t_o \sqsubseteq s_o}_{\text{\upshape in $O\partialto X$}}
  \]
\end{definition}
In other words, when ascending in the order of $\SA$, input transitions can
be added and output transitions can be removed if there is still at least one
output transition afterwards.
  \begin{figure}[t]
    \centering
    \begin{tikzpicture}[x=2cm,y=2cm]
    \begin{scope}[automaton,xshift=-0.38*\textwidth]
      \foreach \statename [count=\count]in {3,1,2} {
        \node[state] (q\statename) at (\count,0) {$\statename$};
      }
      \foreach \statename [count=\count]in {4,6,5} {
        \node[state] (q\statename) at (\count,-1) {$\statename$};
      }
      \foreach \from/\to/\lbl/\nodeargs in
        {1/2/!x/,
        1/3/?a/swap,
        3/4/?a/,
        3/6/!w/,
        2/6/!w/swap,
        4/6/{!x, !y}/swap,
        2/5/?a/,
        5/6/!y/}
      {
        \path[transition] (q\from) edge node[\nodeargs] {\ensuremath{\lbl}} (q\to);
      }
      \path[transition] (q6) edge[loop at=-90] node[alias=loopnode] {\ensuremath{!z}} (q6);
      \node[coalgebra frame,fit={([yshift=1mm]q1.north) (q3) (q5)
      ([yshift=2mm]loopnode.center)}] (frame C) {};
      \node[coalgebra name] at (frame C.north) {$(C,c)$};
    \end{scope}
    \begin{scope}[automaton,xshift=0.38*\textwidth]
      \node[state] (p1) at (0,0) {$1'$};
      \node[state] (p2) at (1,0) {$2'$};
      \node[state] (p6) at (0,-1) {$6'$};
      \node[state] (p5) at (1,-1) {$5'$};
      \foreach \from/\to/\lbl/\nodeargs in
        {1/2/{?a,!x}/,
         2/6/!w/swap,
         2/5/?a/,
         5/6/{?a,!y}/,
         6/1/?a/}
      {
        \path[transition] (p\from) edge node[\nodeargs] {\ensuremath{\lbl}} (p\to);
      }
      \path[transition] (p6) edge[loop at=-90] node[alias=loopnode] {\ensuremath{!z}} (p6);
      \node[coalgebra frame,fit={([yshift=1mm]p1.north) (p1) (p5)
      ([yshift=2mm]loopnode.center)}] (frame D) {};
      \node[coalgebra name] at (frame D.north) {$(D,d)$};
    \end{scope}
    \begin{scope}
      \coordinate (mapsto start) at (frame C.east |- q1.center);
      \path[longmapsto,->] (mapsto start) to
        node {$h$}
        node[alias=substack,below] {}
        (frame D.west |- mapsto start);
        \node[anchor=north] at (substack) {\(
          \begin{array}{c}
            1\mapsto 1'\\
            2\mapsto 2'\\
            3\mapsto 2'\\
            4\mapsto 5'\\
            5\mapsto 5'\\
            6\mapsto 6'
          \end{array}
          \)};
    \end{scope}
    \end{tikzpicture}
    \caption{Examples of suspension automata and a lax $\S$-coalgebra morphism
    between them.}
    \label{figExSA}
  \end{figure}
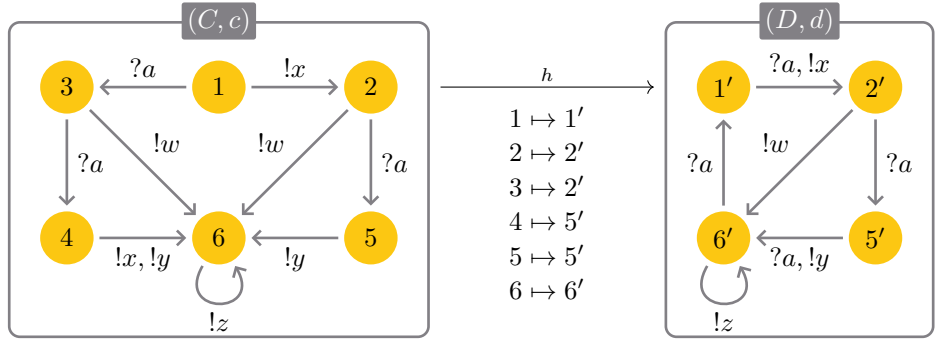
\begin{example}
  We recall two examples of suspension automata from van den Bos et\
  al.~\cite[Fig.~1]{BJM19} in \autoref{figExSA}.
  With the order on $\SA$ (\autoref{defOrderSA}), there is a lax coalgebra
  morphism $h\colon (C,c)\to (D,d)$ between them that identifies some of the
  states: $h(3) = h(2)$ and $h(4) = h(5)$. The map $h$ is only a lax coalgebra
  morphism because there is no input transition for $a$ from $5$ (or $4$) to $6$
  in $(C,c)$, but we have $5'\xtransto{?a} 6'$ in $(D,d)$.
  Conversely, there is an output transition $4\xtransto{!x} 6$ in $(C,c)$ but there is no transition $h(4) = 5' \xtransto{!x} 6' = h(6)$
  in $(D,d)$. 
  Summarizing the above two points, we have:
  \[
    \pi_1(\SA h(c(5))) \sqsubsetneqq \pi_1(d(h(5)))
    \quad\text{and}\quad
    \pi_2(\SA h(c(4))) \sqsupsetneqq \pi_2(d(h(4)))
  \]
\end{example}

The partial order on the functor does not only relax the notion of morphism,
but also gives rise to a new coalgebraic bisimulation notion, which we introduce in \autoref{sec:uncertain-bisim}.

\section{Bisimulation Notions are Coalgebras in Rel}

In this section we recall coalgebraic bisimulation through the lens of relation
liftings; for an extensive introduction see~\cite{jacobs-coalg}. We start by fixing some notation regarding relations.

\begin{notation}
  Given relations $R\subseteq X\times Y$ and $S\subseteq Y\times Z$, the
  composition $R\circ S$ is given by:
  $
    R\circ S :=
    \set{
      (x,z)\in X\times Z\mid
      \exists y\in Y\colon (x,y) \in R\text{ and }(y,z)\in Z
    }
  $. We denote the converse of $R$ by $R^\op = \{(y,x) \mid (x,y) \in R\}$.
  The equality relation (also called the diagonal) on a set $X$ is denoted by
  $\Eq_X = \{(x,x) \mid x\in X\}$.
  Given a map $f \colon X \rightarrow Y$ and a relation $U \subseteq Y \times Y$,
  inverse image is denoted by $(f \times f)^{-1}(U) = \{(x_1,x_2) \mid
  (f(x_1),f(x_2)) \in U\}$.
  The kernel relation of $f$ is given by $\ker f = \{(x_1,x_2) \mid f(x_1) =
  f(x_2)\}$.
  Note that $\ker f = (f \times f)^{-1}(\Eq_Y)$.
  We denote the projections from the product $X\times Y$ by $\pi_1\colon X\times
  Y\to X$ and $\pi_2\colon X\times Y \to Y$.
\end{notation}

For a structural study of (bi)simulation notions on coalgebras, we consider the
fibred category of \emph{relations}:
\begin{definition}
  The category $\Rel$ has objects $(X,R)$, where $X$ is a set and $R\subseteq
  X\times X$, i.e.~$R$ is a relation on $X$. The morphisms $f\colon (X,R)\to
  (Y,S)$ in $\Rel$ are maps $f\colon X\to Y$ that preserve the relation,
  i.e.~$(x_1,x_2)\in R$ implies $(f(x_1), f(x_2))\in S$.
  The obvious forgetful functor is $p\colon \Rel\to \Set$, given by $p(X,R) = X$.
\end{definition}

The forgetful functor $p$ is a fibration; for a thorough introduction, see the
introductory chapter of Jacobs' book~\cite{JacobsCLTT}.
We can express the preservation property of the morphisms in $\Rel$ in a
point-free way: $f \colon X \rightarrow Y$ is a map from $(X,R)$ to $(Y,S)$ in $\Rel$
if and only if
\[
  R \subseteq (f \times f)^{-1}(S).
\]
Equality extends to a functor $\Eq \colon \Set \rightarrow \Rel$,
given by $\Eq(X) = (X,\Eq_X)$ and $\Eq(f) = f$. This is well-defined since we have
$\Eq_X \subseteq (f \times f)^{-1}(\Eq_Y)$ for every map $f \colon X \rightarrow Y$.

To study relations on $F$-coalgebras -- most notably notions of behavioural equivalence inclusion -- we lift the type functor $F$ from $\Set$ to an endofunctor $\hat F$
on $\Rel$.
\[
  \begin{tikzcd}
    \Rel
    \arrow{r}{\hat F}
    \arrow{d}[swap]{p}
    & \Rel
    \arrow{d}{p}
    \\
    \Set
    \arrow{r}{F}
    & \Set
  \end{tikzcd}
\]
For all set functors, such a lifting exists in a canonical
way:

\begin{definition}\label{defRelLift}
  For a functor $F\colon \Set\to\Set$, the relation lifting $\hat F\colon
  \Rel\to\Rel$ is given by
  \twnote{}
  \[
    \hat F (R\subseteq X\times X) = \set{
      (x,y)\in FX\times FX\mid \exists t\in F R\colon
      F\pi_1(t) = x\text{ and }F\pi_2(t) =y
    }.
  \]
\end{definition}
Hence, $\hat F$ transforms relations on $X$ into relations on $FX$.
Before we proceed, we list several standard properties of the relation lifting:

\begin{notheorembrackets}
\begin{lemma}[\cite{jacobs-coalg}]\label{lm:rel-lift}
  For every functor $F\colon \Set\to\Set$, we have
	\begin{enumerate}
		\item Monotonicity: if $R \subseteq S$ then $\hat F(R) \subseteq \hat F(S)$.
		\item Preservation of equality: $\Eq \circ F \subseteq \hat F \circ \Eq$.
		\item Preservation of converse: $\hat F (R^\op) = (\hat F (R))^\op$ for all $R\subseteq X \times X$.
		\item Preservation of inverse images: For a map $f \colon X \rightarrow Y$ and relation $S \subseteq Y \times Y$, we have 
		$$
		\hat F((f \times f)^{-1}(S)) \subseteq (Ff \times Ff)^{-1}(\hat F(S)) \,.
		$$
		Moreover, if $F$ preserves weak pullbacks, then this is an equality.
	\end{enumerate}
\end{lemma}
\end{notheorembrackets}
As a consequence of monotonicity and preservation of inverse images, $\hat{F}$
indeed extends to a lifting of $F$, given on morphisms
by $\hat{F}(f) = F(f)$.

\begin{example} \label{exMealyRelLift}
	In our example of (partial) Mealy machines as coalgebras for $\M X =
  (\set{\question} + O\times X)^I$, a
  relation $R\subseteq X\times X$ is lifted to the 
  relation $\hat \M R \subseteq \M X\times \M X$ given by
  \[
    (s,t) \in \hat\M R \quad\text{ iff }\quad
    \text{for all }i\in I\colon
    \begin{array}[t]{l}
    (s(i) = \question \text{ and } t(i) = \question)\text{ or }
    \\
    (s(i), t(i)) \in \set{
      \big((o,x), (o,y)\big)
      \mid o\in O, (x,y) \in R
    }
    \end{array}
  \]
  Thus, successor structures $s, t \in \M R$ are related by
  $\hat \M R$ if $s$ and $t$ have transitions defined for the same inputs
  $i\in I$, and for all inputs $i\in I$ for which $s(i) = (o,x)$ and $t(i) =
  (o',y)$ are defined,
  both have the same output $o=o'$, and the successor states are related
  $(x,y)\in R$.
\end{example}

The relation lifting is reminiscent of the criterion of a relation $R\subseteq C\times
C$ being a bisimulation. However, we make a distinction between the relation $R$
on the nested elements (or successor states) and the lifted relation on the
structure (or intuitively predecessor states).
If we intuitively let the relation on predecessor and successor states coincide,
then the relation lifting gives rise to bisimilarity as follows~\cite{HermidaJ98,jacobs-coalg}.
\begin{definition}
	A relation $R \subseteq C \times C$ on the state space of a coalgebra $c \colon C \rightarrow FC$
	is a \emph{bisimulation} if $R \subseteq (c \times c)^{-1}(\hat F(R))$.  States
  $x,y\in C$ are called \emph{bisimilar} if there is a bisimulation relating them.
\end{definition}
Note that $(c \times c)^{-1}(\hat F(-)) \colon \Rel_C \rightarrow \Rel_C$
is a monotone map on the complete lattice $\Rel_C = \mathcal{P}(C \times C)$ of relations on $C$, ordered by inclusion.
A bisimulation is thus a post-fixed for this map, and bisimilarity is the greatest post-fixed point,
which is also the greatest fixed point by the Knaster-Tarski theorem. Characterizing bisimilarity
as the greatest fixed point of a monotone map is standard in the classical theory of coinduction~\cite{San:intro12}.

\begin{remark}[Disjoint union of coalgebras]
  In the definition of bisimulation, we consider a relation $R$ on the state
  space of a single coalgebra $c\colon C\to FC$. This bisimulation notion
  straightforwardly generalizes to states of different $F$-coalgebras $x\in C$,
  and $y\in D\xrightarrow{d} FD$, because we can consider the bisimulation
  notion on the disjoint union (i.e.~coproduct) of the coalgebras $(C,c)$ and $(D,d)$:
  \[
    C+D \xrightarrow{~c+d~} FC+FD \xrightarrow{~[F\inl, F\inr]~} F(C+D)
  \]
  where $\inl\colon C\to C+D$ and $\inr\colon D\to C+D$ are the coproduct
  injections and $[\cdots]$ is case distinction (i.e.~the universal mapping property
  of the coproduct).
  So by the bisimilarity of $x$ and $y$ we mean the bisimilarity of $\inl(x)$
  and $\inr(y)$ in the above combined coalgebra. One can easily see that this
  generalization is well-defined: in the special case where $(D,d) := (C,c)$,
  states $x,y$ in $C$ are bisimilar iff $\inl(x), \inr(y)$ are bisimilar in the
  coalgebra on $C+C$.
\end{remark}

\begin{example}
  The relation lifting for $\M$ (\autoref{exMealyRelLift}) thus
  gives rise to the following: a bisimulation on a coalgebra
  $c\colon C\to \M C$ is a relation $R\subseteq C\times C$ such that
  \[
    R \subseteq (c \times c)^{-1}(\hat \M R).
  \]
  Spelling out the inclusion yields that $R$ is a bisimulation iff for all
  $(x,y) \in R$ and $i\in I$:
  \begin{enumerate}
  \item $c(x)(i) = \question$ iff $c(y)(i) = \question$,
  \item if $c(x)(i) = (o,x')\in O\times C$, then $c(y)(i) =
  (o,y')$ for some $y'\in C$ with $(x',y') \in R$, and
  \item if $c(y)(i) = (o,y') \in O\times C$, then $c(x)(i) =
  (o,x')$ for some $x'\in C$ with $(x',y') \in R$.
  \end{enumerate}
  For example, the leaf states $q_1, p_1, p_2, r_1$ in \autoref{figExFunSim} are
  all pairwise bisimilar. However,
  $q_0$ and $p_0$ are not bisimilar: $q_0$ can not mimic the $j$-transition
  of $p_0$. Similarly, $q_0$ and $r_0$ are not bisimilar (and also $p_0$ and $r_0$ are not bisimilar).
\end{example}
In order to still express the compatibility of $q_0$ and $p_0$, we relax the
notion of coalgebraic bisimilarity in the next section.

\section{Uncertain Bisimilarity}\label{sec:uncertain-bisim}
So far, we have not considered the order $\sqsubseteq$ when discussing
bisimulations on coalgebras for a functor $F$. By taking the order into
account, we introduce the notion of \emph{uncertain bisimulation}.
In particular, in the example of Mealy machines it captures
a notion of equivalence where \textqt{unknown} transitions are ignored.
Since we stick to the principle that bisimulation notions are coalgebras in
\Rel, we only need to make use of the order $\sqsubseteq$ when defining a
functor on \Rel. The desired bisimulations will then be coalgebras for this
functor:
\begin{definition}
\label{defUncert}
	The \emph{uncertain relation lifting} of $F$ is defined on a relation $R \subseteq X \times X$ by
	\[
    \uncertOP(R) ~:=~ \mathord{\sqsubseteq_{FX}} \circ \hat F(R) \circ \mathord{\sqsupseteq_{FX}}
  \]
\end{definition}

\begin{remark} \label{hughesJacobsSim}
\autoref{defUncert} is inspired by the notion of
\emph{simulation} on coalgebras by Hughes and Jacobs~\cite{HughesJ04}.
In their work, a simulation
on a coalgebra $(C,c)$ is a relation $R$ such that 
$$
R \subseteq (c \times c)^{-1}(\sqsubseteq_{FC} \circ \mathrel{\hat{F}(R)} \circ \sqsubseteq_{FC})
$$
The lifting $\sqsubseteq \circ \mathrel{\Rel(F)(-)} \circ \sqsupseteq$ of $F$ is referred
to in \emph{op.\ cit.} as the \emph{lax relation lifting}.
\end{remark}
\begin{definition}
  An \emph{uncertain bisimulation} $R$ on a coalgebra $c\colon C\to FC$ is a
  relation $R\subseteq C\times C$ with $R~\subseteq~ (c \times c)^{-1}(\uncertOP(R))$. States
  $x,y\in C$ are called \emph{uncertain bisimilar} if there is an uncertain
  bisimulation relating them. Complementarily, $x,y\in C$ are called \emph{apart} if
  there is no uncertain bisimulation relating them.
\end{definition}
The uncertainty here expresses that in the learning setting, we are not entirely
certain that the two states are bisimilar. With an extension of the system by
a future observation, they might turn out to be non-bisimilar.
With this intuition, the opposite property is simply called \emph{apart}: whenever
two states are separated, they will stay so no matter how the system might be
extended by further transitions.

When unfolding the definitions, we yield the following explicit
characterization:
\begin{lemma}\label{uncertBisimExplicit}
  A relation $R\subseteq C\times C$ on $c\colon C\to FC$ is an uncertain
  bisimulation if and only if for every $(x,y)\in R$ there exists $t\in FR$ such
  that
  \[
    c(x) \sqsubseteq F\pi_1 (t)
    \qquad\text{and}\qquad
    c(y) \sqsubseteq F\pi_2 (t).
  \]
  When representing the witnesses $t$ as a choice function, then we equivalently
  have a map $r\colon R\to FR$ making the projections $\pi_1,\pi_2$ oplax
  coalgebra morphisms $(R,r)\oplaxto (C,c)$:
  \[
    \begin{tikzcd}
      C
      \arrow{d}{c}
      \descto{dr}{\ensuremath{\sqsubseteq}}
      & R
      \descto{dr}{\ensuremath{\sqsupseteq}}
      \arrow{d}{r}
      \arrow{l}[above]{\pi_1}
      \arrow{r}{\pi_2}
      & C
      \arrow{d}{c}
      \\
      FC
      & FR
      \arrow{l}[above]{F\pi_1}
      \arrow{r}{F\pi_2}
      & FC
    \end{tikzcd}
  \]
\end{lemma}\twnote{}
This characterization instantiates to partial Mealy machines as:
\begin{proofappendix}{uncertBisimExplicit}
  Given a relation $R\subseteq C\times C$ on a coalgebra $c\colon C\to FC$, we
  find:
  \begin{enumerate}
  \item[] $R$ is an uncertain bisimulation
  \item[$\Leftrightarrow$] $R\subseteq (c\times c)^{-1}(\uncertOP (R))$
  \item[$\Leftrightarrow$] for all $(x,y) \in R$, we have $(c(x), c(y)) \in
  \uncertOP (R)$
  \item[$\Leftrightarrow$] for all $(x,y) \in R$, we have $(c(x), c(y)) \in
  \mathord{\sqsubseteq_{FC}}\circ \hat F R\circ \mathord{\sqsupseteq_{FC}} $
  \item[$\Leftrightarrow$] for all $(x,y) \in R$, there exists
  $t_x,t_y\in F C$ with
  $c(x) \sqsubseteq t_x$,
  $(t_x,t_y) \in \hat F R$,
  $t_y \sqsupseteq c(y)$
  \item[$\Leftrightarrow$] for all $(x,y) \in R$, there exists
  $(t_x,t_y) \in \hat F R$ with
  $c(x) \sqsubseteq t_x$,
  $c(y) \sqsubseteq t_y$
  \item[$\Leftrightarrow$] for all $(x,y) \in R$, there exists
  $t\in F R$ with
  $c(x) \sqsubseteq F\pi_1 (t)$,
  $c(y) \sqsubseteq F\pi_2 (t)$
  (by \autoref{defRelLift})
  \end{enumerate}
  The last clause is precisely what had to be shown. For the commutativity of
  the diagram, let $r\colon R\to FR$ the map that chooses for each $(x,y)\in R$
  some witness $t\in FR$ with the above stated property. Then, for each $i\in
  \set{1,2}$ and all $(x,y) \in R$, we have
  \[
    F\pi_i(r(x,y))
    \sqsupseteq c(\pi_i(x,y))
  \]
  as desired.
\end{proofappendix}

\begin{lemma}\label{exMealyUncertBisim}%
  For partial Mealy machines $c\colon C\to \M C$ a \emph{reflexive} relation
  $R\subseteq C\times C$ is an uncertain bisimulation
  if and only if for all $(x,y) \in R$ and $i\in I$ we have:
  \begin{equation}
    x\xrightarrow{i/o} x'
    \text{ and }
    y\xrightarrow{i/o'} y'
    \qquad\text{ imply }\qquad
    o=o' \text{ and }
    (x',y')\in R.
    \label{coindMealyUncert}
  \end{equation}
\end{lemma}
  This condition is vacuously satisfied for all $(x,y)\in R$ and $i\in I$
  whenever $x$ or $y$ have no $i$-transition defined.
  In this characterization, we use the mild assumption of $R$ being reflexive in
  order to be able to define the coalgebra structure $r\colon R\to FR$ of
  \autoref{uncertBisimExplicit} in the case where only one of the related states
  has an $i$-transition defined.
  Even without $R$ being reflexive, every uncertain bisimulation $R$
  satisfies \eqref{coindMealyUncert}. Conversely, for every relation $R$
  satisfying \eqref{coindMealyUncert}, the relation $\Eq\cup R$ is an uncertain
  bisimulation (i.e.~we implicitly work with reflexive closure as an up-to technique~\cite{BonchiPPR17}).
\begin{proofappendix}{exMealyUncertBisim}
  We directly verify the general characterization (i.e.~\textqt{up-to
  equality}). Given an arbitrary relation $R\subseteq C\times C$ on $c\colon
  C\to FC$, consider $r\colon R\to \M C$ provided by
  \autoref{uncertBisimExplicit}. For the verification of
  \eqref{coindMealyUncert}, consider $i\in I$, $(x,y) \in R$, and
  transitions
  \[
    x\xrightarrow{i/o} x'
    \text{ and }
    y\xrightarrow{i/o'} y'.
  \]
  By the property of $r$, we have
  \[
    c(x) \sqsubseteq \M \pi_1 (r(x,y))
    \text{ and }
    c(y) \sqsubseteq \M \pi_2 (r(x,y)).
  \]
  Since $c(x)(i) = (o,x')$ and $c(y)(i) = (o', y')$ the inequalities imply that
  $r(x,y)(i) = (o'', p)$, $p\in R$, with
  \[
    o = o'', x' = \pi_1 p,
    o' = o'', y' = \pi_2 p.
  \]
  In other words $o = o'' = o'$ and $(x,y) = p \in R$ as desired.

  For the converse direction, let $R\subseteq C\times C$ be an arbitrary (not
  necessarily reflexive) relation satisfying the coinductive property
  \eqref{coindMealyUncert}. Then, we define an $F$-coalgebra structure
  \[
    r\colon \Eq\cup R\to \M(\Eq\cup R)
  \]
  If $R$ already was
  reflexive, then of course this is a coalgebra on $R$ directly:
  \twnote{}
  \begin{equation}
    r(x,y)(i) = \begin{cases}
      (o,(x',y')) & \text{if }x\xrightarrow{i/o} x'
                    \text{ and }y\xrightarrow{i/o'} y' \\
      (o,(x',x')) & \text{if }x\xrightarrow{i/o} x'
                    \text{ and }c(y)(i) = \question\\
      (o,(y',y')) & \text{if }c(x)(i) = \question
                    \text{ and }y\xrightarrow{i/o} y' \\
      \question & \text{if }c(x)(i) = \question
                    \text{ and }c(y)(i) = \question
    \end{cases}
    \label{proofDefMealyJoin}
  \end{equation}
  In order to see that the image of $r$ indeed restricts to $\M(\Eq\cup R)$,
  note that in the first case, we have that $(x',y')\in R$ by the property
  \eqref{coindMealyUncert} of $R$ (we also have $o=o'$ but we don't need this
  yet). In the second and third case, we use that $(x',x')\in \Eq\cup R$
  resp.~$(y',y') \in \Eq\cup R$.

  For the application of \autoref{uncertBisimExplicit}, we need to verify the
  two inequalities:
  \[
    \begin{tikzcd}
      C
      \arrow{d}{c}
      \descto{dr}{\ensuremath{\sqsubseteq}}
      & \Eq \cup R
      \descto{dr}{\ensuremath{\sqsupseteq}}
      \arrow{d}{r}
      \arrow{l}[above]{\pi_1}
      \arrow{r}{\pi_2}
      & C
      \arrow{d}{c}
      \\
      \M C
      & \M (\Eq\cup R)
      \arrow{l}[above]{\M \pi_1}
      \arrow{r}{\M \pi_2}
      & \M C
    \end{tikzcd}
  \]
  The first inequality
  \[
    c(x) \sqsubseteq \M\pi_1 (r(x,y))
    \quad\text{for all }(x,y) \in \Eq\cup R
  \]
  directly holds, and in the second inequality
  \[
    c(y) \sqsubseteq \M\pi_2 (r(x,y))
    \quad\text{for all }(x,y) \in \Eq\cup R
  \]
  we use that by $R$ satisfying \eqref{coindMealyUncert}, the output of
  $r(x,y)(i)$ matches with the output of $c(y)(i)$ in the first case in the case
  distinction of \eqref{proofDefMealyJoin}.
\end{proofappendix}

	The characterization in \autoref{exMealyUncertBisim} leads to the following coinduction principle:
\begin{proposition}
\label{charMealyUncert}
  States $x,y$ in a partial Mealy machine $c\colon C\to \M C$ are uncertain
  bisimilar iff
  \[
    \text{for all }w\in I^+\colon\qquad
    \semantics{x}(w) \in O
    \text{ and }
    \semantics{y}(w) \in O
    \qquad\Longrightarrow\qquad
    \semantics{x}(w) = \semantics{y}(w)
  \]
\end{proposition}
Dually, $x$ and $y$ are apart iff there is some $w\in I^+$ for which both are
defined but differ: $\question\neq\semantics{x}\neq\semantics{y}\neq\question$.

Recall that the final coalgebra semantics of a state $x\in C$ is a partial map
$\semantics{x}\colon I^+\partialto O$ (in $\nu \M$, \autoref{thmFinalPartialMealy}).
This map sends each input word $w\in I^+$ to
the output symbol of the last transition of the run of $w$, if such a run
exists. If not all required transitions exist, then the map returns $\question$ i.
The characterization in \autoref{charMealyUncert} states that two states are uncertain bisimilar if for all input
words $w\in I^+$, whenever both behaviours $\semantics{x}$, $\semantics{y}$ are
defined, then they must agree.
\begin{proofappendix}{charMealyUncert}
  In the direction $(\Rightarrow)$ follows directly from
  the coinductive rule \eqref{coindMealyUncert}.
  In the converse direction $(\Leftarrow)$, consider states $x,y\in C$ with
  compatible semantics maps $\semantics{x}$, $\semantics{y}$. Define
  the relation
  \[
    R := \set{ (x',y') \mid \exists w\in I^*\colon \text{both $x$ and $y$ have
    runs }x\xtransto{w} x'\text{ and } y\xtransto{w} y'}
  \]
  This relation satisfies the property \eqref{coindMealyUncert}, and so $\Eq\cup
  R$ is an uncertain bisimulation with $(x,y)\in R$ (since $w$ in the
  definition of $R$ ranges over all words).
\end{proofappendix}

\begin{example}\label{exTopOrder}
  If there is a natural transformation $\top\colon 1\to F$ such that $\top_X$ is
  the greatest element of $FX$, then all states in any $F$-coalgebra are
  uncertain bisimilar.
\end{example}
\begin{proofappendix}[Details for]{exTopOrder}
  We show that for any $c\colon C\to FC$, the total relation $R:=C\times C$ is
  an uncertain bisimulation. Define
  $r\colon R\to FR$ by
  \[
    r :\equiv\big(
      R\xrightarrow{!} 1
      \xrightarrow{\top_R} FR
    \big)
  \]
  For both projections $\pi_i\colon R\to C$, $i\in \set{1,2}$, we have:
  \[
    \begin{tikzcd}
      C\times C
      \arrow{d}[swap]{!}
      \arrow{r}{\pi_i}
      \descto{dr}{\(\sqsupseteq\)}
      & C
      \arrow{dd}{c}
      \\
      1
      \arrow{d}[swap]{\top_{C\times C}}
      \arrow{dr}{\top_{C}}
      & {}
      \\
      F(C\times C)
      \arrow{r}[swap]{F\pi_i}
      & FC
    \end{tikzcd}
  \]
  The square commutes laxly because $\top_C$ is the greatest element and the
  triangle commutes because $\top$ is natural.
\end{proofappendix}

\begin{example}\label{exPowOrder}
  For the inclusion order $\subseteq$ on the finitary powerset functor $\Powf$, any two
  states $x,y$ in any $\Powf$-coalgebra $c\colon C\to \Powf C$ are uncertain
  bisimilar.
\end{example}
\begin{proofappendix}[Details for]{exPowOrder}
  Consider $\mathord{\sqsubseteq} := \mathord{\subseteq}$ on $\Powf X$. Given a
  coalgebra $c\colon C\to \Powf C$ we show that the total relation
  $R := C\times C$ is an uncertain bisimilarity. Define the coalgebra structure
  $r\colon C\times C\to \Powf (C\times C)$ by:
  \[
    r(x,y) := \set{ (x',x) \mid x' \in c(x)}
      \cup \set{ (y,y') \mid y' \in c(y)}
  \]
  Then, we need to check that we have oplax coalgebra morphisms
  \[
    \begin{tikzcd}
      C\times C
      \arrow{d}[swap]{r}
      \arrow{r}{\pi_i}
      \descto{dr}{\(\supseteq\)}
      & C
      \arrow{d}{c}
      \\
      \Powf(C\times C)
      \arrow{r}[swap]{\Powf\pi_i}
      & \Powf C
    \end{tikzcd}
    \quad\text{for }i\in\set{1,2}.
  \]
  We verify
  \begin{align*}
    \Powf\pi_1(r(x,y))
    &= \Powf\pi_1(\set{ (x',x) \mid x' \in c(x)}
      \cup \set{ (y,y') \mid y' \in c(y)})
    \\
    &= \set{ x' \mid x' \in c(x)}
      \cup \set{ y \mid y' \in c(y)}
    \\
    &= c(x) \cup \set{ y \mid y' \in c(y)}
    \\ &
    \supseteq c(x)
    = c(\pi_1(x,y))
  \end{align*}
  and
  \begin{align*}
    \Powf\pi_2(r(x,y))
    &= \Powf\pi_2(\set{ (x',x) \mid x' \in c(x)}
      \cup \set{ (y,y') \mid y' \in c(y)})
    \\
    &= \set{ x \mid x' \in c(x)}
      \cup \set{ y' \mid y' \in c(y)})
    \\
    &= \set{ x \mid x' \in c(x)}
      \cup c(y)
      \\ &
    \supseteq c(y)
    = c(\pi_2(x,y))
  \end{align*}
  as desired.
\end{proofappendix}

\begin{example}
  We re-obtain ordinary bisimilarity as the instance where the order
  $\sqsubseteq$ on $FX$ is the discrete poset structure:
  \(
    \mathord{\sqsubseteq}_{FC} :=
    \mathord{\Eq}_{FC}.
  \)
\end{example}

The instance for suspension automata has explicitly been studied in the
literature~\cite{BJM19}:
\begin{notheorembrackets}%
\begin{definition}[{\cite[Def.~15]{BJM19}}] \label{defIocoCompat}
  A relation $R\subseteq C\times C$ on a suspension automaton $c\colon C\to \SA
  C$ is an \emph{(ioco) compatibility relation} if for all $(x,y) \in R$ we have:
  \begin{enumerate}
  \item for all $x\xtransto{?a} x'$ and $y\xtransto{?a} y'$ we have $(x',y') \in R$
  \item there exists $o\in O$ such that $x\xtransto{!o} x'$, $y\xtransto{!o}
  y'$, and $(x',y')\in R$.
  \end{enumerate}
\end{definition}%
\end{notheorembrackets}
\begin{proposition}\label{charIocoCompat}
  A reflexive relation on a suspension automaton is a ioco compatibility
  relation iff it is an uncertain bisimulation.
\end{proposition}
\begin{proofappendix}{charIocoCompat}
  \begin{description}
  \item[$(\Rightarrow)$]
    Consider a ioco compatibility relation $R\subseteq C\times C$ on $c\colon C\to
    \SA C$ with components
    \[
      c =\fpair{c_?,c_!},
      \qquad
      c_?\colon C\to (I\partialto C),
      \text{ and }
      c_!\colon C\to (O\nepartialto C).
    \]
    then we define the required map $r\colon R\to \SA R$ as $r :=
    \fpair{r_?,r_!}$ for the components
    \[
      r_?\colon R\to (I\partialto R)
      \text{ and }
      r_!\colon R\to (O\nepartialto R)
    \]
    The definition of $r_?$ analogous to the proof of \autoref{exMealyUncertBisim}
    \[
      r_?(x,y)(i) = \begin{cases}
        (x',y') & \text{if }x\xtransto{?i} x'\text{ and }y\xtransto{?i} y' \\
        (x',x') & \text{else if }x\xtransto{?i} x' \\
        (y',y') & \text{else if }.\xtransto{?i} .' \\
        \question & \text{otherwise}.
      \end{cases}
    \]
    \[
      r_!(x,y)(o) = \begin{cases}
        (x',y') & \text{if }x\xtransto{!o} x'\text{ and }y\xtransto{!o} y' \\
        \question & \text{otherwise}.
      \end{cases}
    \]
    It is now straightforward to verify that
    \[
      \begin{tikzcd}[sep=10mm]
        R
        \arrow{r}{\pi_k}
        \arrow{d}{r_?}
        \descto{dr}{\(\sqsupseteq\)}
        & C
        \arrow{d}{c_?}
        \\
        (I\partialto R)
        \arrow{r}{I\partialto \pi_k}
        & (I\partialto C)
      \end{tikzcd}
      \text{ and }
      \begin{tikzcd}[sep=10mm]
        R
        \arrow{r}{\pi_k}
        \arrow{d}{r_!}
        \descto{dr}{\(\sqsubseteq\)}
        & C
        \arrow{d}{c_!}
        \\
        (I\nepartialto R)
        \arrow{r}{I\partialto \pi_k}
        & (I\nepartialto C)
      \end{tikzcd}
      \text{ for all }k\in\set{1,2}.
    \]
    So in total each projection is an oplax morphism $\pi_k\colon (R,r)\oplaxto
    (C,c)$.

  \item[$(\Leftarrow)$]
    Given a coalgebra $r\colon R\to \SA R$, again writing $r =\fpair{r_?,r_!}$,
    whose projections are oplax morphisms $\pi_k\colon (R,r)\oplaxto
    (C,c)$, for $k\in \set{1,2}$. We verify that $R$ is a (ioco) compatibility
    relation:
    \begin{enumerate}
    \item Like in the proof of \autoref{exMealyUncertBisim}, the condition on
    inputs is satisfied: whenever we have $x\xtransto{?a} x'$ and
    $y\xtransto{?a} y'$, since $r\colon R\to \SA R$ is above both, there must be
    a transition for input $a$ in $r_?$ showing that $(x',y')\in R$.

    \item For $(x,y)\in R$, the non-empty partial map $r_!\colon R\to
    (O\nepartialto R)$ provides us with some $o\in O$ and $(x',y')\in R$ such that
    \(
      r_!(x,y)(o) = (x',y')
    \). It remains to show that
    both $x$ and $y$ also have a transition for the output $o$.
    Here, we use that the projections are oplax morphisms and verify:
    \begin{description}
    \item[] $\pi_k\colon (R,r) \oplaxto (C,c)$ for $k\in \set{1,2}$
    \item[$\Longrightarrow$]
      $\SA\pi_k(r(x,y)) \sqsupseteq c(\pi_k(x,y))$ for $k\in \set{1,2}$.
    \item[$\Longrightarrow$]
      $(O\nepartialto \pi_k)(r_!(x,y)) \sqsubseteq c_!(\pi_k(x,y))$ for $k\in
      \set{1,2}$ (by \autoref{defOrderSA}).
    \item[$\Longrightarrow$]
      $(O\nepartialto \pi_1)(r_!(x,y)) \sqsubseteq c_!(x)$
      and $(O\nepartialto \pi_2)(r_!(x,y)) \sqsubseteq c_!(y)$
    \item[$\Longrightarrow$]
      $\pi_1(r_!(x,y)(o)) = c_1(x)(o)$
      and $\pi_2(r_!(x,y)(o)) = c_2(x)(o)$
      (for $o\in O$ from above for which $r_!(x,y)(o)$ is defined)
    \item[$\Longrightarrow$]
      $x' = c_1(x)(o)$
      and $y' = c_2(x)(o)$
    \item[$\Longrightarrow$]
      $x\xtransto{!o} x'$
      and
      $y\xtransto{!o} y'$
    \end{description}
    as desired.

    \qedhere
    \end{enumerate}
  \end{description}
\end{proofappendix}

In the proof it is relevant that the output transitions of suspension automata
are non-empty partial maps $C\to (O\nepartialto C)$. Non-emptiness means that
whenever there is a coalgebra structure $r\colon R\to \SA R$ on a relation
$R\subseteq C\times C$, then all related states $(x,y)\in R$ have at least one common
output $o\in O$. This is reflected by the existentially quantified
condition in the definition of ioco compatibility.

\subsection{Properties}

Having discussed instances, we now uniformly establish general properties of
uncertain bisimilarity.
We start by listing properties of uncertain relation lifting, analogous to \autoref{lm:rel-lift}.

\begin{lemma}\label{lm:rel-lift-lax}
  For any functor $F_\Pos \colon \Set\to\Pos$, we have the following properties of uncertain relation lifting:
	\begin{enumerate}
		\item Monotonicity: if $R \subseteq S$ then $\uncertOP(R) \subseteq \uncertOP (S)$.
		\item Preservation of equality: $\Eq \circ F \subseteq \uncertOP \circ \Eq$.
		\item Preservation of converse: $\uncertOP (R^\op) = (\uncertOP (R))^\op$ for all $R\subseteq X \times X$.
		\item\label{item:pres-inv-lax} Preservation of inverse images: For a map $f \colon X \rightarrow Y$ and relation $S \subseteq Y \times Y$, we have 
		$$
		\uncertOP((f \times f)^{-1}(S)) \subseteq (Ff \times Ff)^{-1}(\uncertOP(S)) \,.
		$$
	\end{enumerate}
\end{lemma}
\begin{proofappendix}{lm:rel-lift-lax}
\begin{enumerate}
	\item Monotonicity: follows immediately from monotonicity of $\hat F$.
	\item Preservation of equality: 
	\begin{align*}
	\Eq_{FX} &\subseteq \mathord{\sqsubseteq_{FX}} \circ \sqsupseteq_{FX} \tag{reflexivity}\\
		&= \mathord{\sqsubseteq_{FX}} \circ \Eq_{FX} \circ \sqsupseteq_{FX} \tag{\autoref{lm:rel-lift}} \\
		&= \mathord{\sqsubseteq_{FX}} \circ \hat{F}(\Eq_X) \circ \sqsupseteq_{FX}
	\end{align*}
	\item Preservation of converse:
	\begin{align*}
			(\sqsubseteq_{FC} \circ \hat F (R) \circ \sqsupseteq_{FC})^\op
			&= (\sqsupseteq_{FC})^\op \circ (\hat F(R))^\op \circ (\sqsubseteq_{FC})^\op \\
			&= \mathord{\sqsubseteq_{FC}} \circ \hat F(R^\op) \circ \sqsupseteq_{FC} \tag{\autoref{lm:rel-lift}}
	\end{align*}

	\item Preservation of inverse images:
	\begin{align*}
			 &\uncertOP((f \times f)^{-1}(R))  \\
			&= \mathord{\sqsubseteq_{FX}} \circ \hat{F}((f \times f)^{-1}(R)) \circ \sqsupseteq_{FX} \\
			&\subseteq (Ff \times Ff)^{-1}(\sqsubseteq_{FY}) \circ \hat{F}((f \times f)^{-1}(R)) \circ (Ff \times Ff)^{-1}(\sqsupseteq_{FY}) \tag{$Ff$ monotone} \\
			&\subseteq (Ff \times Ff)^{-1}(\sqsubseteq_{FY}) \circ (Ff \times Ff)^{-1}(\hat{F}(R)) \circ (Ff \times Ff)^{-1}(\sqsupseteq_{FY}) \tag{\autoref{lm:rel-lift}} \\
			& \subseteq (Ff \times Ff)^{-1}(\sqsubseteq_{FY} \circ \hat{F}(R) \circ \sqsupseteq_{FY}) \\
			& = (Ff \times Ff)^{-1}(\uncertOP(R)) \,.
	\end{align*}
	
\end{enumerate}
\end{proofappendix}
Similar to the case of $\hat{F}$, by monotonicity and preservation of inverse images, $\uncertOP$ extends to a lifting of $F$.
Uncertain bisimilarity is reflexive and
symmetric:
\begin{lemma}\label{uncertReflSym}
	The equality relation $\Eq_C$ on any coalgebra $(C,c)$ is an uncertain bisimulation,
	and if $R \subseteq C \times C$ is an uncertain bisimulation then so is $R^\op$.
\end{lemma}
\begin{proofappendix}{uncertReflSym}
	For reflexivity:
	$$
		\Eq_C 
			\subseteq (c \times c)^{-1}(\Eq_{FC})
			\subseteq (c \times c)^{-1}(\uncertOP(\Eq_C))
	$$
	by \autoref{lm:rel-lift-lax}.
	
	For symmetry:
	\begin{align*}
		R^\op
			&\subseteq ((c \times c)^{-1}(\uncertOP(R)))^\op \tag{$R$ an uncertain bisimulation, $(-)^{\op}$ monotone} \\
			&= (c \times c)^{-1}(\uncertOP(R)^\op) \\
			&= (c \times c)^{-1}(\uncertOP(R^\op))
	\end{align*}
	by \autoref{lm:rel-lift-lax}.
\end{proofappendix}
Unsurprisingly, uncertain bisimilarity is not transitive: even though two states
$x$ and $z$ are certainly non-bisimilar (i.e.~not uncertain bisimilar), there
can be a state $y$ that is uncertain bisimilar to both $x$ and $z$ (e.g.~$p_0$,
$q_0$, $s_0$ in \autoref{figNotionComparison}).
Similarly, ioco compatibility is known to not be transitive in
general~\cite[Ex.~17]{BJM19}.

This lack of transitivity makes it non-trivial to characterize the uncertain
bisimilarity in terms of being identifiable by a morphism, the way holds for
normal bisimilarity.
Still, we can show some preservation results that match the intuition that
the order $\sqsubseteq$ adds transitions (or other information).
Since uncertain bisimilarity of two states means that there is no conflict
in their existing transition behaviour, they stay uncertain bisimilar if we omit
transitions:
\begin{lemma} \label{oplaxPreservesUncertain}
  Uncertain bisimilarity is preserved by oplax morphisms: whenever states $x,y$
  in $(C,c)$ are uncertain bisimilar, then for every oplax coalgebra morphism
  $h\colon (C,c)\to (D,d)$, the states $h(x)$ and $h(y)$ are uncertain bisimilar
  in $(D,d)$.
\end{lemma}
\begin{proofappendix}{oplaxPreservesUncertain}
  Let $x,y$ in $c\colon C\to FC$ be related by an uncertain bisimulation
  $R$ witnessed by $r\colon R\to FR$ (\autoref{uncertBisimExplicit}) making the
  projections oplax morphisms $\pi_1,\pi_2\colon (R,r)\oplaxto (C,c)$. The
  projections together form an injective map:
  \[
    \fpair{\pi_1,\pi_2}\colon R\longrightarrow C\times C.
  \]
  For an oplax coalgebra morphism $h\colon (C,c)\oplaxto (D,d)$,
  consider the image factorization of the composed map $\fpair{\pi_1,\pi_2}\circ
  (h\times h)$:
  \begin{equation}
    \begin{tikzcd}
      R
      \arrow{r}{e}
      \arrow{d}[swap]{\fpair{\pi_1,\pi_2}}
      &
      S
      \arrow{d}{\fpair{\pi_1',\pi_2'}}
      \\
      C\times C
      \arrow{r}{h\times h}
      & D\times D
    \end{tikzcd}
    \label{eqDiagImS}
  \end{equation}
  For disambiguation, we use $\pi_1',\pi_2'$ for the projections of the
  resulting image $S\subseteq D\times D$. Working with sets (and the axiom of
  choice), let $a\colon S\to R$ be a splitting of the surjective map $e\colon
  R\to S$, that is we have
  \[
    e\circ a = \id_S.
  \]
  We now verify that the coalgebra structure
  \[
    s :\equiv\big(
      S\xrightarrow{a} R
      \xrightarrow{r} FR
      \xrightarrow{Fe} FS
    \big)
  \]
  makes $\pi_i'$ an oplax coalgebra morphism $(S,s)\to (D,d)$ for all $i\in
  \set{1,2}$. To this end, take the diagram:
  \[
    \begin{tikzcd}
      S
      \arrow{d}[swap]{a}
      \arrow{r}{\id_S}
      \arrow[shiftarr={xshift=-16mm}]{ddd}[swap]{s}
      \descto[xshift=-8mm]{ddd}{Def.~\(s\)}
      & S
      \arrow{rd}{\pi_i'}
      \descto{d}{\eqref{eqDiagImS}}
      \\
      R
      \arrow{d}[swap]{r}
      \arrow{ur}{e}
      \arrow{r}{\pi_i}
      \descto{dr}{\ensuremath{\sqsupseteq}}
      & C
      \arrow{r}{h}
      \arrow{d}{c}
      \descto{dr}{\ensuremath{\sqsupseteq}}
      & D
      \arrow{d}{d}
      \\
      FR
      \arrow{d}[swap]{Fe}
      \arrow{r}[swap]{F\pi_i}
      & FC
      \arrow{r}[swap]{Fh}
      \descto{d}{\eqref{eqDiagImS}}
      & FD
      \\
      FS
      \arrow[rounded corners, to path={-| (\tikztotarget)
      \tikztonodes}]{rru}[swap,pos=0.25]{F\pi_i'}
      & {}
    \end{tikzcd}
    \qquad\text{for all }i\in \set{1,2}.
  \]
  Using the functoriality of $F_\Pos$ means that we can compose the (laxly)
  commuting cells and that in total, we obtain $F\pi_i'(s(p)) \sqsupseteq
  d(\pi_i'(p))$ for all $p\in S$, as desired.
\end{proofappendix}

Conversely, we can show that if two states can be identified by a lax coalgebra
morphisms, then they are uncertain bisimilar.
For the corresponding proof for (canonical) relation liftings $\hat F$, one uses
weak pullback preservation as a sufficient condition for preservation of inverse images.
For uncertain bisimilarity we will simply make preservation of inverse images an assumption, referred to as
\emph{stability}. This terminology follows Hughes and Jacobs~\cite{HughesJ04},
who define a similar condition for their lax relation lifting.

\begin{definition}
	The functor $F_\Pos$ is called \emph{stable} if $\uncertOP$ commutes with
	inverse images on reflexive relations, i.e.~the inclusion in Item~\ref{item:pres-inv-lax} of \autoref{lm:rel-lift-lax} is an equality if $S$ is reflexive.
\end{definition}
\begin{remark}\label{invImgCommut}
	Contrary to the variant in~\cite{HughesJ04}, we require the converse of \autoref{lm:rel-lift-lax}.\ref{item:pres-inv-lax}
	only for \emph{reflexive} relations. The reason is that even for the case of Mealy machines, $F=\M$, the converse of \autoref{lm:rel-lift-lax}.\ref{item:pres-inv-lax} does not hold if we drop that assumption.
\end{remark}
\begin{proofappendix}{invImgCommut}
  For the partial Mealy functor $F=\M$, consider the inclusion of sets
  \[
  \begin{tikzcd}[column sep=10mm]
    X=\set{x}
    \arrow[hook]{r}{f}
    &
    Y=\set{x,y}
  \end{tikzcd}
  \]
  and the relation $S := \set{(x,y)}$ on $Y$.
  Fix inputs $I=1$ and outputs $O=1$, so $\M$ boils down to
  \[
    \M X \cong \set{\question} + X.
  \]
  This induces the following relations:
  \begin{align*}
    (f\times f)^{-1}(S) &= \emptyset
    \\
    \hat \M((f\times f)^{-1}(S)) &= \set{(\question,\question)}
    \\
    \hat \M_\sqsubseteq((f\times f)^{-1}(S)) &=
    (\mathord{\sqsubseteq_{\M X}} \circ \hat \M (f\times f)^{-1}(S)
    \circ \mathord{\sqsupseteq_{\M X}})
    =
    \set{(\question,\question)}
  \end{align*}
  But on the other hand, we have:
  \begin{align*}
    \hat \M S
    &= \set{ (\question,\question), (x,y) }
    \\
    \hat\M_\sqsubseteq S
    &= (\mathord{\sqsubseteq_{\M Y}} \circ \hat \M S \circ
    \mathord{\sqsupseteq_{\M Y}})
    = \set{ (\question,\question), (x,y), (\question,y), (x,\question) }
    \\
    (\M f\times \M f)^{-1}(\hat\M_\sqsubseteq S)
    &= \set{ (\question,\question), (x,\question) }
  \end{align*}
  Thus, $\hat \M_\sqsubseteq((f\times f)^{-1}(S)) \not\supseteq (\M f\times \M
  f)^{-1}(\hat\M_\sqsubseteq S)$.
\end{proofappendix}
\begin{example}\label{exMealyStable}
  $\hat M_\sqsubseteq$ is stable.
  \twnote{}
\end{example}
\begin{proofappendix}{exMealyStable}
  The verification is reminiscent of the proof of \autoref{exMealyUncertBisim}.
  For $f\colon X\to Y$ let $S\subseteq Y\times Y$ be reflexive.
  For the verification of the converse of
  \autoref{lm:rel-lift-lax}.\ref{item:pres-inv-lax}, consider
  \[
    (t,s) \in (\M f\times \M f)^{-1}(\sqsubseteq_{FY}\circ \hat \M(S)\circ
    \sqsupseteq_{FY})
  \]
  Define $w\in \M (X\times X)$ by
  \[
    w \quad:= \quad i\mapsto
    \begin{cases}
      (o_1,(x_1,x_2))
      &\text{if }t(i) = (o_1,x_1)
      \text{ and }s(i) = (o_2,x_2)
      \\
      (o_1,(x_1,x_1))
      &\text{if }t(i) = (o_1,x_1)
      \text{ and }s(i) = \question
      \\
      (o_2,(x_2,x_2))
      &\text{if }t(i) = \question
      \text{ and }s(i) = (o_2,x_2)
      \\
      \question
      &\text{if }t(i) = \question
      \text{ and }s(i) = \question.
    \end{cases}
  \]
  We clearly have $t \sqsubseteq \M \pi_1 (w)$. Since $(t,s)$ are related by
  above lifting, we have $o_1=o_2$ in the first case and so also $s \sqsubseteq
  \M \pi_2 (w)$. Moreover, we have that
  \[
    \M(f\times f)(w) \in \M(S) \subseteq \M(Y)
  \]
  because $(x_1,x_2)\in S$ (in the first case) and because $S$ is reflexive (in the
  other cases).
  Using that $\M$ preserves inverse images, we obtain
  \[
    w\in \M((f\times f)^{-1}(S)).
  \]
  By definition of $\hat \M$, this implies that
  \[
    (\M\pi_1(w),\M\pi_2(w)) \in \M((f\times f)^{-1}(S))
  \]
  which are the witnesses for
  \[
    (t,s) \in \mathord{\sqsubseteq_{FX}} \circ \M((f\times f)^{-1}(S)) \circ \mathord{\sqsupseteq_{FX}}
    = \hat \M_\sqsubseteq ((f\times f)^{-1}(S)),
  \]
  we had to be shown.
\end{proofappendix}

\begin{lemma} \label{reflectUncert}
	Suppose that $F_\Pos$ is stable. Then any lax coalgebra morphism 
	$h\colon (C,c)\laxto (D,d)$ reflects uncertain bisimilarity, that is,
	if $R \subseteq C \times C$ is a reflexive uncertain bisimulation relation then so
	is $(h \times h)^{-1}(R)$.
\end{lemma}
\begin{proofappendix}{reflectUncert}
    First observe that, for all $x,y \in C$, if $d(h(x)) \mathrel{\uncertOP(R)} d(h(y))$, then, since $h$ is a lax coalgebra morphism, we get
    $Fh(c(x)) \sqsubseteq_{FD} d(h(x)) \mathrel{\uncertOP(R)} d(h(y)) \sqsupseteq_{FD} Fh(c(y))$, and hence
	$Fh(c(x)) \mathrel{\uncertOP(R)} Fh(c(y))$ by transitivity of $\sqsubseteq_{FD}$.
    That is, we have:
	\begin{equation}\label{eq:laxcoalg-upperbound}
		(h \times h)^{-1} \circ (d \times d)^{-1}(\uncertOP(R)) \subseteq (c \times c)^{-1} \circ (Fh \times Fh)^{-1}(\uncertOP(R)) \,.
	\end{equation}
	We proceed as follows:
	\begin{align*}
			& (h \times h)^{-1}(R) \\
			&\subseteq (h \times h)^{-1} \circ (d \times d)^{-1}(\uncertOP(R)) \tag{$R$ an uncertain bisimulation}\\
			&\subseteq (c \times c)^{-1} \circ (Fh \times Fh)^{-1}(\uncertOP(R)) \tag{by~\eqref{eq:laxcoalg-upperbound}} \\
			&= (c \times c)^{-1} \circ \uncertOP ((h \times h)^{-1}(R)) \tag{$F_\Pos$ is stable}
	\end{align*}
	Hence $(h \times h)^{-1}(R)$ is an uncertain bisimulation.
\end{proofappendix}
As a consequence, under the assumption of stability,
if states $x,y$ of a coalgebra are identified by a lax homomorphism $h$ then they are
uncertain bisimilar.
\begin{corollary} \label{kerIsUncert}
	Suppose that $F_\Pos$ is stable.
  Then the kernel $\ker h$ of a lax coalgebra morphism $h\colon (C,c)\laxto (D,d)$
  is an uncertain bisimulation. 
\end{corollary}
\begin{proofappendix}{kerIsUncert}
	Follows from \autoref{reflectUncert} and the fact that the equality relation is a bisimulation (\autoref{uncertReflSym}).
\end{proofappendix}

This gives half a characterization theorem of uncertain bisimilarity
(assuming stability):
\begin{equation}
  \text{\begin{tabular}{c}
    States $x,y$ can be identified \\
    by a lax coalgebra morphism
  \end{tabular}}
  \Longrightarrow
  \text{\begin{tabular}{c}
    States $x,y$ are \\
    uncertain bisimilar
  \end{tabular}}
  \label{eqFunSimToBisim}
\end{equation}
For standard bisimilarity, the converse direction also holds: whenever states
$x,y$ are bisimilar (in the ordinary sense), then they can be identified by an
(ordinary) coalgebra morphism.
For uncertain bisimilarity however, the converse direction even fails when
restricting to tree-shaped Mealy machines:

\begin{example}\label{exCounter}
  Consider the partial Mealy machine $\M$ in
  \autoref{figTreeCounterExample} for $I=\set{v,w,i}$ and $O= \set{a,b,o}$.
  \begin{figure}%
  \begin{minipage}[b]{.5\textwidth}%
  \centering%
  \begin{tikzpicture}[automaton,y=16mm,x=16mm]
    \node[state] (p) at (0,0) {$p$};
    \node[state] (x) at (1,0) {$x$};
    \node[state] (xp) at (2,0) {$x'$};

    \node[state] (q) at (0,-1) {$q$};
    \node[state] (y) at (1,-1) {$y$};

    \node[state] (qp) at (0,-2) {$q'$};
    \node[state] (z) at (1,-2) {$z$};
    \node[state] (zp) at (2,-2) {$z'$};
    \path[transition]
    (p) edge node {$w/o$} (x)
    (q) edge node {$w/o$} (y)
    (p) edge node[swap] {$v/o$} (q)
    (q) edge node[swap] {$v/o$} (qp)
    (qp) edge node {$w/o$} (z)
    (x) edge node {$i/a$} (xp)
    (z) edge node {$i/b$} (zp)
    ;
  \end{tikzpicture}
  \caption{Mealy machine of \autoref{exCounter} $\ldots$}
  \label{figTreeCounterExample}
  \end{minipage}%
  \hfill%
  \begin{minipage}[b]{.48\textwidth}
  \[
    \begin{array}{@{\semantics{p}(}l@{)\,}l@{\qquad\semantics{q}(}l@{)\,}l}
      w & = o & w & = o \\
      wi & = a & wi & = \question \\
      v & = o & v & = o \\
      vw & = o & vw & = o\\
      vv & = o & vv & =\question \\
      vvw & = o & vvw & =\question \\
      vvwi & = b & vvwi  & = \question \\
      vwi & = \question  & vwi & = b\\
    \end{array}
  \]
  \caption{$\ldots$ and its semantics}
  \label{figTreeCounterExampleSemantics}
  \end{minipage}%
  \end{figure}%
  In this machine, $p$ and $q$ are uncertain bisimilar, because their semantics
  matches on all defined input words, as verified in
  \autoref{figTreeCounterExampleSemantics} (using \autoref{charMealyUncert}).
  However, there is no lax coalgebra morphism $f\colon \M \laxto \mealyN $ with $f(p)
  = f(q)$. To see this, first observe that for any such $f$, we can derive the following equalities:
  \[
    f(p) \xrightarrow{w/o} f(x)
    \quad\text{and}\quad
    f(q) \xrightarrow{w/o} f(y)
    \quad
    \text{implies}
    \quad
    f(x) = f(y)\,,
  \]
  \[
    f(p) \xrightarrow{v/o} f(q) \xrightarrow{w/o} f(y)
    \quad\text{and}\quad
    f(q) \xrightarrow{v/o} f(q') \xrightarrow{w/o} f(z)
    \quad
    \text{implies}
    \quad
    f(y) = f(z)
  \]
  and hence $f(x) = f(z)$. By \autoref{kerIsUncert} this means $x$ and $z$
  are uncertain bisimilar; but $i$ witnesses that $x$ and $z$ are apart -- a contradiction!
  Thus, there is no $f\colon \M\to\mealyN$ with $f(p)=f(q)$.
\end{example}

\subsection{Characterization via Simulations}

We can remedy the failure of the converse direction of \eqref{eqFunSimToBisim} by going from functional simulations (i.e.~lax coalgebra
morphisms) to proper simulations in the sense of spans of (lax) morphisms.

There are multiple ways to define simulations between coalgebras for functors
$F\colon \Set\to\Set$ equipped with order $\sqsubseteq$.
The way that Hughes and Jacobs~\cite{HughesJ04} define simulations (see also
\autoref{hughesJacobsSim}) between coalgebras $(C,c)$ and $(D,d)$ corresponds to
a relation $R\subseteq C\times D$ and a structure $r\colon R\to FR$
making the projections oplax and lax morphisms:
\[
  \begin{array}{ll}
  \pi_1\colon (R,r)\oplaxto (C,c) \\
  \pi_2\colon (R,r)\laxto (D,d)
  \end{array}
  \text{ that is, diagrammatically: }
    \begin{tikzcd}
      C
      \arrow{d}{c}
      \descto{dr}{\ensuremath{\sqsubseteq}}
      & R
      \descto{dr}{\ensuremath{\sqsubseteq}}
      \arrow{d}{r}
      \arrow{l}[above]{\pi_1}
      \arrow{r}{\pi_2}
      & D
      \arrow{d}{d}
      \\
      FC
      & FR
      \arrow{l}[above]{F\pi_1}
      \arrow{r}{F\pi_2}
      & FD
    \end{tikzcd}
\]
Note that due to using both lax and oplax morphisms, such a simulation is
subtly different from the diagram in \autoref{uncertBisimExplicit}.
We can now show that this span-based definition of simulation characterizes
uncertain bisimilarity:

\begin{proposition}\label{characterHJSim}
  Given that $F_\Pos$ is stable, the following are equivalent for all states
  $x,y$ in a coalgebra $c\colon C\to FC$:
  \begin{enumerate}
  \item $x$ and $y$ are uncertain bisimilar.
  \item There is a state $z\in D$ in another coalgebra $(D,d)$ and a simulation
  $S\subseteq C\times D$ in the style of Hughes and Jacobs such that $(x,z)\in
  S$ and $(y,z)\in S$.
  \end{enumerate}
\end{proposition}

The second item intuitively means that the states $x$ and $y$ can be
\textqt{identified} by a simulation. We obtained a converse to the implication in
\eqref{eqFunSimToBisim} when replacing \textqt{lax coalgebra morphism} with
\textqt{simulation}.
In the proof of the first direction (top to bottom), we use that in sets, every
surjective function $e\colon X\to Y$ has a right-inverse $a\colon Y\to X$ (i.e.~with
$e\circ a = \id_Y$), using the axiom of choice. In the second direction (bottom
to top), we use the stability of $F_\Pos$.
\begin{proofappendix}{characterHJSim}
  \begin{description}
  \item[($\Rightarrow$)]
    Let $r\colon R\to FR$ be a witnessing uncertain bisimulation (by
    \autoref{uncertBisimExplicit}) with $(x,y)\in R$.
    By diagram chasing, we directly have:
    \[
      \begin{tikzcd}[column sep=25mm]
        C
        \arrow{dd}[swap]{c}
        \descto{dr}{\(\sqsupseteq\)}
        & R+R
        \descto{dr}{\(\commutes\)}
        \arrow{l}[swap]{[\pi_1,\pi_2]}
        \arrow{r}{[\id_R,\id_R]}
        \arrow{d}{r+r}
        & R
        \arrow{dd}{r}
        \\
        & FR + FR
        \arrow{d}{[F\inl,F\inr]}
        \arrow{dl}[sloped,above]{[F\pi_1,F\pi_2]}
        \arrow{dr}[sloped,above]{[F\id_R,F\id_R]}
        & {}
        \\
        FC
        & F(R+R)
        \arrow{l}{F[\pi_1,\pi_2]}
        \arrow{r}[swap]{F[\id_R,\id_R]}
        & FR
      \end{tikzcd}
    \]
    The desired $z \in R$ is simply the pair $(x,y)\in R$, because $R+R$ relates
    $x$ with $z$ and also $y$ with $z$:
    \[
      \begin{tikzcd}[row sep=0pt,column sep=14mm]
      \inl(x,y)
      \arrow[mapsto]{r}{[\pi_1,\pi_2]}
      & x
      \\
      \inr(x,y)
      \arrow[mapsto]{r}{[\pi_1,\pi_2]}
      & y
      \end{tikzcd}
      \qquad\text{ and }\qquad
      \begin{tikzcd}[row sep=0pt,column sep=14mm]
      \inl(x,y)
      \arrow[mapsto]{r}{[\id_R,\id_R]}
      & (x,y) = z
      \\
      \inr(x,y)
      \arrow[mapsto]{r}{[\id_R,\id_R]}
      & (x,y) = z
      \end{tikzcd}
    \]
    This does not yet prove the claim because the span at hand is not jointly
    injective in general. Thus, define a relation $S\subseteq C\times (R+R)$ as
    the image factorization of the map
    \[
      R+R\xrightarrow{[\id_R,\id_R]} R
    \]
    that is
    \[
      \begin{tikzcd}[column sep=14mm]
        R+R
        \arrow[shiftarr={yshift=6mm}]{rr}{[\id_R,\id_R]}
        \arrow{r}{e}
        & S
        \arrow{r}{\fpair{\pi_1',\pi_2'}}
        & C\times R
      \end{tikzcd}
    \]
    Choose a right-inverse of $e$, that is, some map $a\colon S\to R+R$ with
    $e\circ a = \id_S$. This now shows that the projections $\pi_1'$ and
    $\pi_2'$ of $S$ have the desired properties:
    \[
      \begin{tikzcd}
        && S
        \arrow{d}{a}
        \arrow{dl}[swap]{\id_S}
        \arrow{drr}{\pi_2'}
        \\
        C
        \arrow{dd}{c}
        & S
        \arrow{l}[swap]{\pi_1'}
        & R+R
        \arrow{d}
        \arrow{l}[swap]{e}
        \arrow[bend left=20]{ll}[below]{[\pi_1,\pi_2]}
        \arrow{rr}[pos=0.3]{[\id_R,\id_R]}
        \descto{drr}{\commutes}
        & & R
        \arrow{dd}{r}
        \\[5mm]
        {}
        \descto[yshift=3mm]{rr}{\(\sqsupseteq\)}
        & & F(R+R)
        \arrow{dll}[sloped,above]{F[\pi_1,\pi_2]}
        \arrow{d}{Fe}
        \arrow{drr}[sloped,above]{F[\id_R,\id_R]}
        & & {}
        \\
        FC
        && FS
        \arrow{ll}{\pi_1'}
        \arrow{rr}[swap]{\pi_2'}
        && FR
      \end{tikzcd}
    \]
    Here, the coalgebra structure $R+R$ and the (lax) commutativity of the two
    big cells are taken from the diagram before.
    The above verification that $z$ \textqt{identifies} $x$ and $y$
    straightforwardly adapts to $(x,z)\in S$ and $(y,z) \in S$.
  \item[($\Leftarrow$)]
    Consider a simulation $S\subseteq C\times D$, that is, a structure $s\colon
    S\to FS$ making the projections oplax and lax morphisms:
    \[
      \pi_1\colon (S,s)\oplaxto (C,c)
      \qquad\text{and}\qquad
      \pi_2\colon (S,s)\oplaxto (D,d)
    \]
    Moreover, assume some $z\in D$ with $(x,z)\in S$ and $(y,z) \in S$. Since
    $\pi_2(x,z) = z = \pi_2(y,z)$, \autoref{kerIsUncert} shows that $(x,z)$ and
    $(y,z)$ are uncertain bisimilar in the coalgebra $(S,s)$ (here, we use that $F_\Pos$ is
    stable). Since oplax morphisms preserve uncertain bisimilarity
    (\autoref{oplaxPreservesUncertain}), we obtain that
    $x = \pi_1(x,z)$ and $y = \pi_2(y,z)$ are uncertain bisimilar in the
    coalgebra $(C,c)$.
    \qedhere
  \end{description}
\end{proofappendix}

Another slightly different notion of simulation on coalgebras arises from the
approach to bisimilarity via open maps~\cite{joyal96,WDKH2019}. Here, a
simulation between $(C,c)$ and $(D,d)$ is again a relation $R\subseteq
C\times D$ equipped with a coalgebra structure $r\colon R\to FR$ such that
\begin{enumerate}
  \item the projection $\pi_1$ is a coalgebra morphism $\pi_1\colon (R,r)\to (C,c)$, and
  \item the projection $\pi_2$ is a lax coalgebra morphism $\pi_2\colon
  (R,r)\laxto (D,d)$:
\end{enumerate}
\[
    \begin{tikzcd}
      C
      \arrow{d}{c}
      \descto{dr}{\ensuremath{\commutes}}
      & R
      \descto{dr}{\ensuremath{\sqsubseteq}}
      \arrow{d}{r}
      \arrow{l}[above]{\pi_1}
      \arrow{r}{\pi_2}
      & D
      \arrow{d}{d}
      \\
      FC
      & FR
      \arrow{l}[above]{F\pi_1}
      \arrow{r}{F\pi_2}
      & FD
    \end{tikzcd}
\]
Hence, any open-map-style simulation is also a simulation in the style of Hughes
and Jacobs. In our leading examples, the converse inclusion also holds. To this
end, we impose another assumption on the functor:
\begin{definition}\label{defRestricting}
  We call the order $\sqsubseteq$ on the functor $F$ \emph{restricting} if
  for all maps $f\colon X\to Y$ and all $s\in FX$, $t\in FY$
  we have
  \begin{equation}
    t\sqsubseteq_{FY} Ff(s)
    \qquad
    \Longrightarrow
    \qquad
    \text{there is some }
    s'\sqsubseteq s
    \text{ with }
    t = Ff(s').
    \label{eqRestricting}
  \end{equation}
\end{definition}
The idea behind $s'$ is that it is the restriction of $s$ to those transitions
that are defined in $t$, so that $Ff\colon FX\to FY$ maps $s'$ to $t$:
\begin{example}\label{exMealyRestricting}
  The functor $\M$ for partial Mealy machines is restricting:
  for $f\colon X\to Y$, $s\in \M X$, and $t\sqsubseteq_{\M Y} \M f(s)$, define
  \[
    s'\in \M X = (\set{\question} + O\times X)^I
    \quad\text{ by }\quad
    s'(i) = \begin{cases}
      \question &\text{if }t(i) = \question \\
      s(i)    &\text{otherwise}.
    \end{cases}
  \]
  This definition makes $s'\sqsubseteq s$ true because for all $i\in I$,
  whenever $s'(i)$ is defined (i.e.~$s'(i)\neq\question$) then $s(i)$ is
  defined, too. The inequality $t\sqsubseteq_{\M Y} \M f(s)$ implies
  \[
    (s'(i) = \question
    \quad\Longleftrightarrow\quad
    t(i) = \question)
    \qquad\text{for all }i\in I
  \]
  and moreover, whenever $s'(i) = (o,x)$ for $i\in I$, then $t(i) = s(i) =
  s'(i)$. Hence, $\M f(s') = t$.
\end{example}

\begin{remark}
  In the definition of \emph{restricting}, we have $t\sqsubseteq Ff (s)$ as the
  condition and
  then construct some $s'$ with $s'\sqsubseteq s$. Thus, one
  might be tempted to think that there is a Galois connection hidden. Note
  however, that this is not the case in the example of partial Mealy machines
  because the construction of $s'$ does depend on $s$! \jrnote{} Hence, it is not
  possible to construct an adjoint map $FY \to FX$ in general.
\end{remark}

\begin{lemma}
  \label{coalgebraRestriction}
  If $F$ is restricting, then
  for every oplax morphism $h\colon (C,c)\oplaxto (D,d)$, there is a structure
  $c'\colon C\to FC$ such that $c'(x) \sqsubseteq c(x)$ for all $x\in C$ making
  $h$ a (proper) coalgebra morphism.
  \[
    \begin{tikzcd}
      C
      \arrow{d}{c}
      \arrow{r}{h}
      \descto{dr}{\ensuremath{\sqsupseteq}}
      & D
      \arrow{d}{d}
      \\
      FC
      \arrow{r}{Fh}
      & FD
    \end{tikzcd}
    \qquad
    \Longrightarrow
    \qquad
    \exists c'\colon
    \qquad
    \begin{tikzcd}
      C
      \arrow[bend right]{d}[left]{c}
      \arrow[bend left]{d}[right]{c'}
      \arrow{r}{h}
      \descto{d}{\ensuremath{\sqsupseteq}}
      \descto[xshift=2pt]{dr}{\commutes}
      & D
      \arrow{d}{d}
      \\
      FC
      \arrow{r}{Fh}
      & FD
    \end{tikzcd}
  \]
\end{lemma}
\begin{proofappendix}{coalgebraRestriction}
  For each $x\in C$, we have $d(h(x)) \sqsubseteq Fh(c(x))$. So for each $x\in
  C$, apply that $F$ is restricting (for $s := c(x)$, $t:=d(h(x))$), and denote
  the resulting $s'\in FC$ by $c'(x)$. This defines a map $c'\colon C\to FC$
  with $c'(x) \sqsubseteq c(x)$ (for all $x\in C$) and $Fh(c'(x)) = d(h(x))$.
\end{proofappendix}
This lemma turns Hughes/Jacobs simulations into simulations
in the style of open maps:
\begin{lemma} \label{equivalentSimStyle}
  Given that $\sqsubseteq$ is restricting, the following are equivalent for any
  relation $S\subseteq C\times D$ on coalgebras $(C,c)$, $(D,d)$:
  \begin{enumerate}
  \item there is a map $S\to FS$ making $S$ a simulation in the style of Hughes and Jacobs.
  \item there is a map $S\to FS$ making $S$ a simulation in the style used in open maps.
  \end{enumerate}
\end{lemma}
\begin{proofappendix}{equivalentSimStyle}
  Since every coalgebra morphism is also an oplax morphism, the direction
  $(\Leftarrow)$ is clear. For the other direction $(\Rightarrow)$, consider a
  structure $s\colon S\to FS$ such that $\pi_1\colon (S,s)\oplaxto (C,c)$ is an
  oplax morphism and $\pi_2\colon (S,s)\laxto (D,d)$ is a lax morphism.
  Apply \autoref{coalgebraRestriction} on $\pi_1$ yielding the desired map
  $s'\colon S\to FS$:
  \[
    \begin{tikzcd}[sep=12mm]
      C
      \arrow{d}[swap]{c}
      & S
      \arrow{l}[swap]{\pi_1}
      \arrow{r}{\pi_2}
      \arrow[bend right]{d}[swap]{s'}
      \arrow[bend left]{d}{s}
      \descto{d}{\(\sqsubseteq\)}
      \descto{dr}{\(\sqsubseteq\)}
      \descto{dl}{\commutes}
      & D
      \arrow{d}[swap]{d}
      \\
      FC
      & FS
      \arrow{l}[swap]{F\pi_1}
      \arrow{r}{F\pi_2}
      & FD
    \end{tikzcd}
  \]
\end{proofappendix}

Thus, we can combine \autoref{equivalentSimStyle} and the previous
characterization \autoref{characterHJSim}:
\begin{theorem}\label{thmTotalSim}
  Given that $F_\Pos$ is stable and that $\sqsubseteq$ is restricting, the
  following are equivalent for all states
  $x,y$ in a coalgebra $c\colon C\to FC$:
  \begin{enumerate}
  \item $x$ and $y$ are uncertain bisimilar.
  \item There is a state $z\in D$ in another coalgebra $(D,d)$ and a simulation
  $S\subseteq C\times D$ in the style of Hughes and Jacobs such that $(x,z)\in
  S$ and $(y,z)\in S$.
  \item There is a state $z\in D$ in another coalgebra $(D,d)$ and an
  open-map-style simulation
  $S\subseteq C\times D$ such that $(x,z)\in
  S$ and $(y,z)\in S$.
  \end{enumerate}
\end{theorem}

\begin{example}
  For partial Mealy machines, the abstract definitions of simulation instantiate
  to the usual notion of simulation between $(C,c)$ and $(D,d)$ when considering
  the Mealy machines as deterministic LTSs for the alphabet $I\times O$: a
  simulation is a relation $R\subseteq C\times D$ such that for all $(x,z)\in R$, whenever 
  $x\xtransto{i/o} x'$ then there is some $z'\in D$ such that $z\xtransto{i/o}
  z'$ and $(x',z')\in R$.
  The characterization in \autoref{characterHJSim} shows that
  for all states $x,y \in C$ in $c\colon C\to\M C$ we have
\begin{equation*}
  \text{\begin{tabular}{c}
    $x$ and $y$ are \\
    uncertain bisimilar
  \end{tabular}}
  \Longleftrightarrow
  \text{\begin{tabular}{c}
    There is a state $z$ in some $d\colon D\to \M D$\\
    such that $z$ simulates $x$ and $y$
  \end{tabular}}
\end{equation*}
\end{example}

\begin{example}
  For the compatibility relation on suspension automata, a similar equivalence
  holds. In the specific simulation notion (called
  \emph{coinductive ioco relation} \cite[Def.~4]{BJM19}), the input transitions
  are preserved in the usual direction and the output transitions are preserved
  in the converse direction. Then, it is shown that states $x,y$ in a suspension
  automaton are compatible iff there is a state $z$ in another suspension
  automaton which conforms (according to the \emph{ioco} relation) to both $x$ and $y$~\cite[Lem.~24.2]{BJM19}.
  \twnote{}
\end{example}

\twnote{}
\takeout{}

\section{Conclusions and Future Work}%
\twnote{}

We introduced \emph{uncertain bisimilarity}, a notion to talk about behavioural
compatibility on a coalgebraic level of generality. Instances include both partial 
Mealy machines and the \emph{ioco} conformance relation from model-based testing.
The setting is tailored
towards the lack of knowledge in automata learning games. We are optimistic that
this generalization provides a step from the $\lsharp$ learning
algorithm~\cite{VGRW22} towards new coalgebra learning algorithms. While previous categorical 
frameworks~\cite{BarloccoKR19,UrbatS20,Heerdt20,ColcombetPS21} generalize Angluin's classical $L^*$ algorithm, the development of a variant
of $\lsharp$ at a high level of generality could be useful, as the experiments \cite{VGRW22}
point to a better performance in the case of Mealy machines.

So far, we have shown that uncertain bisimilarity is equivalent to being 
simulated by a same state. A similar observation might be
lifted to the final coalgebra by defining a suitable simulation order on the
final coalgebra. In this context it would be interesting to explicitly connect our
results to the similarity quotients in~\cite{Levy11}.

Standard coalgebraic bisimilarity can also be characterized as indistinguishability
via formulas of coalgebraic modal logic~\cite{Schroder08,KupkeP11}.
We are confident that uncertain bisimilarity can be characterized in similar
terms. Since it is a coinductively defined relation involving a non-standard relation lifting, a good starting point may
be the framework in~\cite{KupkeR21}, although that does not provide a canonical construction of a logic but only the infrastructure
for proving expressiveness and adequacy.
To obtain such a logic, it should be feasible to transfer modalities from an existing system type functor (e.g.~$\M_T$) to the functors 
involving order and partial behaviours (e.g.~$\M$), such that properties like adequacy and or even expressiveness are
inherited. These distinguishing modal formulas can then serve as witnesses for
disproving uncertain bisimilarity, that is, for showing apartness.

The relation lifting based definition of uncertain bisimilarity makes it amenable to the use of coalgebraic up-to techniques as developed in~\cite{BonchiPPR17}. This could be helpful in the development of efficient algorithms for checking uncertain bisimilarity, which could be particularly
interesting to check compatibility of ioco specifications as studied in~\cite{BJM19}.
\jrnote{}

\label{maintextend}
\bibliographystyle{plain}
\bibliography{refs}
\vfill

\newpage
\ifthenelse{\boolean{proofsinappendix}}{%
\section{Omitted Proofs and Further Details}
\closeoutputstream{proofstream}
\input{\jobname-proofs.out}
}{%
}

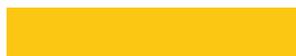
\end{document}